\documentclass[a4paper, 11pt]{article}
\usepackage{amsmath,comment}
\usepackage{graphicx}
\usepackage{epsfig}
\usepackage{epstopdf}
\usepackage{latexsym}
\usepackage{amssymb} % or amsfonts
\usepackage{scalerel,stackengine}
\usepackage{xcolor,url}
\stackMath
\newcommand\reallywidehat[1]{%
\savestack{\tmpbox}{\stretchto{%
  \scaleto{%
    \scalerel*[\widthof{\ensuremath{#1}}]{\kern-.6pt\bigwedge\kern-.6pt}%
    {\rule[-\textheight/2]{1ex}{\textheight}}%WIDTH-LIMITED BIG WEDGE
  }{\textheight}%
}{0.5ex}}%
\stackon[1pt]{#1}{\tmpbox}%
}
\renewcommand{\Bbb}{\mathbb}

\def\b{\begin{eqnarray}}
\def\e{\end{eqnarray}}

\begin{document}
\title{Point vortex dynamics influenced by the surface motion}
\author{Delia Ionescu-Kruse\footnote{Simion Stoilow Institute of Mathematics of the Romanian Academy,
Research Unit No. 6,
 P.O. Box 1-764, RO-014700, Bucharest, Romania;
e-mail: Delia.Ionescu@imar.ro},
        Rossen I. Ivanov\footnote{School of Mathematics and Statistics, Technological University Dublin, Grangegorman Lower,
Dublin D07 ADY7, Ireland;
e-mail: rossen.ivanov@tudublin.ie};
Michail D. Todorov \footnote{Institute of Mathematics and Informatics, Bulgarian Academy of Sciences,  8 Acad. Georgi Bonchev Str.,  1113 Sofia, Bulgaria;
mdtodorov@math.bas.bg}
}
\maketitle

\begin{abstract}

We study a coupled system of differential equations, which models the dynamics of fluid with a free surface and a point vortex in the body of the fluid.
For long surface waves of small amplitude, the effects of the interaction between the waves and the point vortex are modeled within the framework of the Boussinesq regime. 
It turns out that the solitary waves on the surface are not destroyed by the interaction with the vortex, and, as a matter of fact, the solitary waves remain practically unaffected for a significant range of the vortex strength. As a result, a further simplification of the model is proposed, where the vortex motion under solitons propagating on the surface is determined from a system of decoupled ordinary differential equations.

\end{abstract}

\section{Introduction}

We consider a two-dimensional, inviscid and incompressible fluid, with the fluid domain bounded below by a horizontal fixed bottom and above by a free surface. Without restrictions on the vorticity field, we provided very recently \cite{DIK-RI}, a general description of the intricate interaction between surface motion and the motion in the bulk of the fluid. The governing equations are mathematically complicated and they cannot be split into separate sets of equations for the free surface and for the bulk of the fluid. For the chosen surface and volume variables, these nonlinear equations can be written as evolutionary equations, and they are expressed with the aid of the Dirichlet-Neumann operator and the Green function of the Laplace operator in the fluid domain.
Many of the previous studies have either completely neglected the vorticity or, in order to isolate the effect of background vorticity and to avoid mathematical complexity, considered the simpler case of constant vorticity
(see, for example, \cite{CI}-\cite{CIP},   \cite{Craig&Groves},  \cite{IK-M}, \cite{Wahlen},  \cite{Zakharov68} and the references therein).
A variational formulation for water waves with vertical vorticity has been proposed in \cite{CoBo}.
In the case of an arbitrary vorticity, we also  mention \cite{LMMR}  where a non-canonical Poisson bracket has been introduced and it was shown that the equations of motion for incompressible flows with a free boundary having surface tension are Hamiltonian relative to this structure. These Poisson brackets have the awkwardness of non-closure in the class of admissible functionals.

Another class of problems with vorticity in the fluid body is that of vortices that interact with the free surface of the fluid. There is no universally accepted approach for the investigation of these problems. For example, in \cite{Ty,Ty2} the motion of a single vortex and vortex pair near a free surface is studied and the related initial/boundary-value problem is solved in terms of a Taylor expansion in time. 
A system of differential-integral equations is derived in \cite{CKa}, which describes the surface - point vortex interactions in a shallow water regime for an arbitrary number of submerged vortices. The fluid domain is periodic in the horizontal direction. Then a long-time numerical simulations are performed which capture the nonlinear interactions between the surface and the vortices. In \cite{CKa2} a numerical method for the simulation of the simultaneous evolution of a free-surface wave and an existing vortex patch is developed. In \cite{Fish,Will} experiments with vortex pairs and surface waves are reported.

The general framework presented in \cite{DIK-RI} is convenient for the study of the nonlinear interactions between the surface and the bulk of the fluid in general. In particular, it can be applied to nonlinear free-surface flows over a point vortex immersed in the fluid domain. We employ this framework and focus on the Boussinesq and KdV propagation regimes, characterized by the small-amplitude and long-wave assumptions and by the balanced between nonlinearity and dispersion. By assuming that the point vortex does not move under the action of its own field, we obtain a model for the point vortex dynamics, coupled to the motion of the free surface over a flat bottom.
In these regimes, the appropriate Green function for the point-vortex problem is well approximated by the Green function on an infinite strip; we use its expression from \cite{Melnikov} in our further analysis. In addition, we use the formal series expansion of the Dirichlet-Neumann operators. 

In Sections 2-3 we provide a brief summary of the KdV-type model of surface motion coupled to the vortex dynamics. In Section 4, we investigate the model under the assumption that the surface waves are not affected by the vortex. We analyze the vortex motion under the exact KdV one-soliton and two-soliton waves (for the surface velocity) which travel unperturbed on the surface. Thus, the velocity field in the fluid body is produced entirely by the soliton on the surface. Our findings show that in this simplified setup, in which practically the point vortex is reduced to a ``regular fluid particle", the trajectories are bounded in space. For a detailed study of the motion of a fluid particle in the velocity field of the long-wave surface motion modeled by the KdV equation, see also \cite{Bor}.  Moreover, the shape of the trajectories depends on the depth at which the point vortex is located initially. The extent of the bounded horizontal motion of the vortex is estimated analytically. The presented figures reflect the vortex trajectories for different values of the initial positions of the vortex. The plots are produced by Mathematica.

In Section 5 we present the numerical investigations of the model. They show that the decoupling of the surface dynamics described by the KdV equation is justifiable, since we observe qualitatively the same behavior: the point vortex moves within a bounded interval and the surface radiation waves created by the vortex do not affect significantly the propagation of the KdV soliton. The figures presented reflect the dynamics obtained numerically from the coupled equations for various values of the vortex strength as well as for several different values of the soliton parameters.

\section{Equations of motion for two-dimensional water waves with arbitrary vorticity}

\begin{figure}[h!]
\begin{center}
\fbox{\includegraphics[totalheight=0.3\textheight]{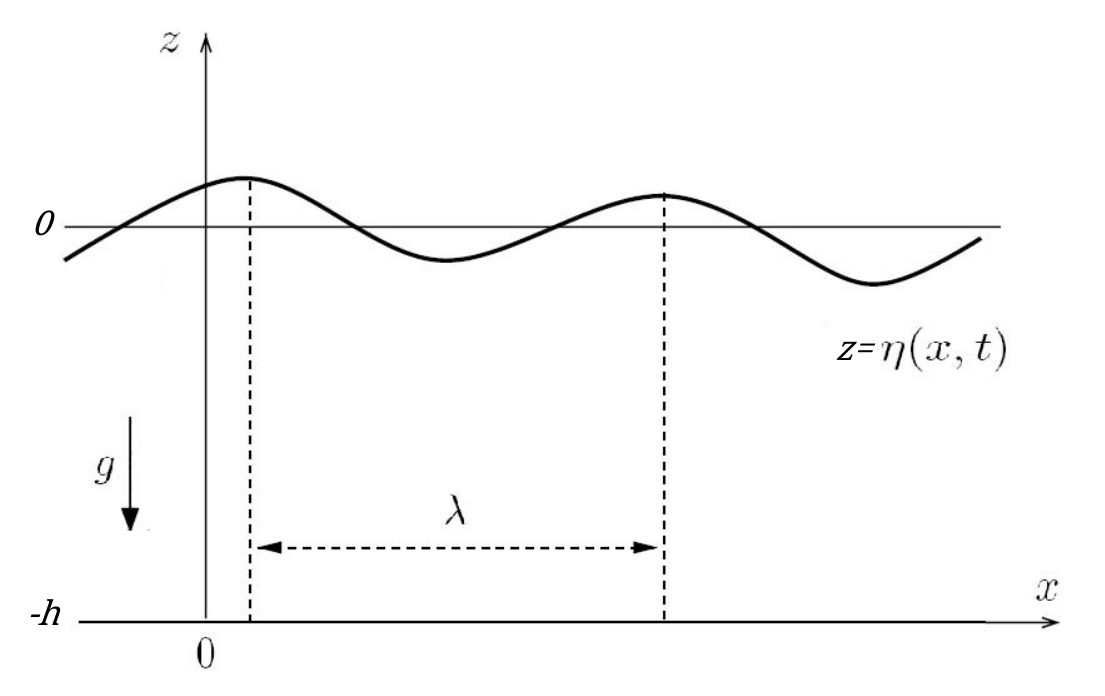}}
\caption{The fluid domain.}
\label{fig:setup}
\end{center}
\end{figure}

The motion of incompressible and inviscid fluids in the Euclidian plane domain
$$\Omega_\eta\ =\ \{(x,z)\in\Bbb{R}^2:\  -h\ <\ z\ <\ \eta(x,t)\},$$
 bounded below by a flat rigid bottom
$$B\ =\ \{(x,z)\in\Bbb{R}^2:\ z\ =\ -h\}$$
and above by the free surface
$$S_\eta\ =\ \{(x,z)\in\Bbb{R}^2:\ z\ =\ \eta(x,t)\},$$
  see Fig. \ref{fig:setup},
is governed by  the following boundary value problem:
\begin{equation}
\begin{split}
u_t\ +\ uu_x\ +\ vu_z\ =\ &  -\ p_x \\
 v_t\ +\ uv_x\ +\ vv_z\ =\  & -\ p_z-\ g \quad \textrm{in} \quad \Omega_{\eta}
 \label{Euler}\\
u_x\ +\ v_z\ =\ &  0 \quad \qquad \qquad\,  \textrm{in} \quad \Omega_{\eta} \\
p\ =\ &  p_{\textrm{atm}} \,  \qquad \quad \, \textrm{ on }\quad  S_\eta\\
 v\ =\ & \eta_t\ +\ u\eta_x \quad \textrm{ on }\quad S_\eta\\
 v\ =\ & 0 \, \,  \qquad \qquad\,\,\,  \textrm { on }\quad  B,
 \end{split}
\end{equation}
  where $(u(x,z,t), v(x,z,t))$ is the velocity field of the
water - no motion takes place in the $y$-direction, $p(x,z,t)$
denotes the pressure, $g$ is the constant gravitational
acceleration, $p_{\textrm{atm}}$ being  the constant atmospheric pressure. We set the constant water density $\rho=1$.

\noindent We define  the scalar vorticity of the two-dimensional  flow by
\begin{align} \omega(x,z,t)\ =\ u_z\ -\  v_x\quad \textrm{ in } \quad \Omega_{\eta}.
\label{vorticity}\end{align}
 We now turn to a brief consideration of the results that can be obtained when a non-constant vorticity field $\omega(x,z,t)$ is introduced into Euler's equation.\\
Taking the curl of the  Euler equations in  \eqref{Euler}, we get the evolution of the vorticity in time
  \begin{align}
\omega_t\ +\ u\, \omega_x\ +\ v\, \omega_z\ =\ 0 \quad \textrm{ in } \quad \Omega_{\eta}.\label{omega_t}
\end{align}
For two-dimensional flows, the incompressibility equation in  \eqref{Euler} ensures the existence
of a stream function $\psi(x, z, t)$ - determined up to an additive term that depends
only on time -  by
\begin{align}
u\ =\ \psi_z, \quad v\ =\ -\psi_x.\label{psi}
\end{align}
We assume that the fluid is at rest at $x\rightarrow \pm \infty$, i.e., a shear current is not present in the domain $\Omega_\eta.$
According to Helmholtz-Hodge decomposition (see, for example,  \cite{BPB}, \cite{Ebin&Marsden}  and the references therein), the velocity field
decomposes uniquely as
\begin{equation}\label{eq1}
  \begin{split}
     u\ =\ & \varphi_x\ +\ r_z\ \stackrel{\eqref{psi}}{=}\ \psi_z \\
      v\ =\ & \varphi_z\ -\ r_x\ \stackrel{\eqref{psi}}{=}\ -\psi_x
  \end{split}
\end{equation} where $\varphi(x,z,t)$ is the velocity potential, and $(r_z,-r_x)$ is a divergence free vector field, which is tangent to the free surface $S_{\eta}$.
Thus, from \eqref{vorticity}-\eqref{eq1}, the nonlinear  Euler's equations and the incompressibility equation
are replaced by  the following equations:
\begin{align}\label{evol-omega}
\omega_t\ =\  \psi_x \omega_z\ -\ \psi_z\omega_x \quad \textrm{ in }\quad  \Omega_\eta,
\end{align}
\begin{align}
\Delta \varphi\ =\ 0 \quad & \textrm{ in }\quad  \Omega_\eta\label{Laplace-varphi} \\
\Delta r\ =\  \omega \quad & \textrm{ in }\quad \Omega_\eta.\label{rGreen}
\end{align}
We now turn to a careful formulation of the  boundary conditions that define the water-wave problem \eqref{Euler},
in terms of the functions  $\varphi(x,z,t)$, $r(x,z,t)$,  $\psi(x,z,t)$ and the free surface $\eta(x,t)$.  We concentrate on waves which decay fast  as $|x| \to \infty$.
The kinematic boundary conditions in \eqref{Euler}, at the free surface and at the flat rigid bottom,  become  (see  \cite {DIK-RI}):
\begin{align}
\eta_t\ =\ - \psi_x\Big|_S\ -\ \psi_z\Big|_S\eta_x\ =\ \varphi_z\Big|_S\ -\ \varphi_x\Big|_S\eta_x\   \quad & \textrm{ on }\quad S_\eta\label{KBC1}\\
 r_x\ +\ r_z\eta_x\ =\ 0 \quad & \textrm{ on }\quad S_\eta\label{KBC2}\\
 r\ =\ 0 \quad & \textrm{ on }\quad S_\eta\label{KBC3}\\
r\ =\ 0 \quad & \textrm{ on }\quad B \label{KBC4}\\
 \varphi_z\ =\ 0 \quad & \textrm{ on }\quad B.\label{KBC5}
\end{align}
In the fluid domain,  the pressure is given by
 the following generalization of  Bernoulli's equation  (see \cite {DIK-RI}):
\begin{align*}
\varphi_t\ +\ \frac{1}{2}|\nabla\psi|^2\ +\  p\ +\ gz\ +\ \partial_x^{-1}\left[\ r_{zt}\ -\  \omega\, \psi_x\right] \ =\ \mathcal{C}(t),
\end{align*}
where $\mathcal{C}(t)$ is an arbitrary function of $t$.
By  appropriate choice of  the function $\mathcal{C}(t)$,
the dynamic boundary condition to be satisfied at the free surface is
\begin{align}\label{bernoulli}
(\varphi_t)_S\ +\ \frac{1}{2}|\nabla\psi|_S^2\  +\ g\eta\ +\
\big[\partial_x^{-1}\left[\ r_{zt}\ -\  \omega\, \psi_x\right]\big]_S
=\ 0 \quad \textrm{ on }\quad S_\eta,
\end{align}
the subindex $S$ being used for the evaluations at the free surface.

Thus, we get the system \eqref{Euler} in the form \eqref{evol-omega}-\eqref{bernoulli}, in terms of  the vorticity $\omega(x,z,t)$  and the fast-decaying functions $\varphi(x,z,t)$, $r(x,z,t)$, $\psi(x,z,t)$ and $\eta(x,t)$. Furthermore, with the aid of the Dirichlet-Neumann operator and the Green function of the Laplace operator in the fluid domain, the problem \eqref{evol-omega}-\eqref{bernoulli} can be
recast, see \cite{DIK-RI},
in an evolutionary form of a system of PDEs with time derivatives separated from other terms, the evolution  variables  being $\omega$,  $r$, $\eta$ and $\xi$, where $\xi$ is
the
restriction of $\varphi$ to the free surface:
\begin{align}
\xi(x,t)\ :=\ \varphi(x,\eta(x,t),t) \quad \textrm{on} \quad S_{\eta}.\label{xi}
\end{align}
 The Dirichlet-Neumann operator  $G$ associated to the domain $\Omega_\eta$, is given by (see \cite{Craig})
\begin{align}\label{G}
G \xi :=\ \sqrt{1+\eta_x^2}\ (\nabla\varphi)_S \cdot \mathbf{n},
\end{align}
$\mathbf{n}$ being the unit outward normal to the free surface,
${\bf n}=\frac{1}{\sqrt{1+\eta_x^2}}(-\eta_x, 1)$.

The solutions  $\varphi$ to  the boundary value problem \eqref{Laplace-varphi}, \eqref{KBC5}, \eqref{xi}, regular in the strip $-h<z<H$, $H$ = const being the upper bound of $\eta$, can be written as (see \cite{DIK-RI})
\begin{align}\label{varphi}
\varphi(x,z)\ =\ \cosh\big((z+h)D\big)\big[:\cosh\big((\eta+h)D\big):\big]^{-1}\ \xi,
\end{align}
where the operator $D$ is defined by
\begin{align}\label{D}
\begin{split}
D=- i\, \partial _x,\quad \partial_x= i\, D,\quad  \partial_x^{-1}= -i\, D^{-1}, \quad D^{-1}=i\, \partial_x^{-1},
\end{split}
\end{align} and
 the  other operators\footnote{The action of an operator $\mathcal{G}(D)$ on functions $f(x)=\frac{1}{2\pi}\int e^{ikx} \hat{f}(k) dk $ is given by the integral $$ \mathcal{G}(D) f(x)=\frac{1}{2\pi}
\int e^{ikx} \mathcal{G}(k)\hat{f}(k) dk . $$} by
$$
\cosh\big((z+h)D\big)\ =\  1\ +\ \frac{(z+h)^2D^2}{2!}\ + \ \frac{(z+h)^4D^4}{4!}\ +\ ...
$$
 $:\cosh\big((\eta+h)D\big):$ being the normal ordering of the operator $\cosh\big((\eta+h)D\big)$,  that is, in its expression
 \begin{align}
:\cosh\big((\eta+h)D\big):\ =\  1\ +\ \frac{(\eta+h)^2D^2}{2!}\ + \ \frac{(\eta+h)^4D^4}{4!}\ +\ ... ,
\end{align}
first are the powers of the function $\eta(x,t)+h$  which depends on $x$ and then the powers of the derivative operator  $D$.

In order to solve the other boundary value problem \eqref{rGreen}, \eqref{KBC3}-\eqref{KBC4}, we use 
the Green function of the Laplace operator in the fluid domain,
that is, the solution  $\Gamma(\mathbf{x}, \mathbf{x}_0)$ to  the problem
\begin{align}\label{Gamma_omega}
\begin{split}
\Delta \Gamma\ &=\ \delta(\mathbf{x}-\mathbf{x}_0)  \quad  \textrm{ in }\quad \Omega_\eta \\
\Gamma\ &=\ 0 \quad \quad\qquad\,\,\,  \textrm{ on } \quad  B\cup S_\eta,
\end{split}
\end{align}
with  $\mathbf{x}=(x,z)\in\Omega_\eta$, $\mathbf{x}_0=(x_0,z_0)$ a  fixed point    in $\Omega_\eta$ and  $\delta$  the Dirac delta function. For the  expression  of this function, see the formula (43) in \cite{DIK-RI}.
Then, by  Green's  representation formula (see, for example,  \cite{Haberman}), the solution to the problem \eqref{rGreen}, \eqref{KBC3}-\eqref{KBC4},  can be written as
\begin{align}\label{r}
r(\mathbf{x})\ =\  \int_{\Omega_\eta} \Gamma(\mathbf{x}, \mathbf{x_0})\ \omega(\mathbf{x_0})\ d\mathbf{x}_0.
\end{align}
From \eqref{eq1},  \eqref{varphi}, \eqref{D} and \eqref{r},
 we obtain for the stream function  (see \cite{DIK-RI}):
\begin{align}\label{50}
\psi(x,z)\
=\  &  \int_{\Omega_\eta}  \Gamma(\mathbf{x},\mathbf{x}_0)\ \omega(\mathbf{x_0})\ d\mathbf{x}_0\nonumber\\
&+i\   \sinh\big((z+h)D\big) \, \left[:\cosh\big((\eta+h)D\big):\right]^{-1}\xi,
\end{align}
where  the operator 
$$
\sinh\big((z+h)D\big)\ =\  (z+h)D\ +\ \frac{(z+h)^3D^3}{3!}\ + \ \frac{(z+h)^5D^5}{5!}\ +\ ...
$$

In view of the boundary conditions  \eqref{KBC1}, \eqref{KBC2},  and the formula \eqref{50}, the Dirichlet-Neumann  operator  \eqref{G} has the following explicit expression in terms of the free surface and the flat bottom  (see \cite{DIK-RI}):
\begin{align}\label{G1}
G \xi\ =&\  -\eta_x\xi_x\ +\ (1+\eta_x^2)\, : \sinh\big((\eta+h)D\big): D\, \left[: \cosh\big((\eta+h)D\big):\right]^{-1} \xi.
\end{align}
By taking into account \eqref{KBC1}, \eqref{G}  and \eqref{varphi},
the Bernoulli relation \eqref{bernoulli} on the free surface yields
 the evolution equation for $\xi$ defined in \eqref{xi}.\\
Finally, the system of evolution equations  for the variables $\omega$, $r$, $\eta$ and $\xi$,  can be written as
 (see \cite{DIK-RI}):
\begin{align}\label{eqs1}
\begin{split}
\omega_t\ =&\  \psi_x \omega_z\ -\ \psi_z\omega_x \qquad \qquad\qquad \hspace{4.3cm} \textrm{ in }\,\,  \Omega_\eta\\
r_t\ =&\ \mathcal{R}(x,z) \hspace{7.8cm}  \textrm{ in }\,\,  \Omega_\eta\\
\eta_t\ =& \ G\xi \hspace{8.7cm} \textrm{on} \,\, S_{\eta}\\
\xi_t\ =&\   \frac{\eta_x\,  \xi_x}{1+\eta_x^2}\,  G\xi  +\  \frac{1}{1+\eta_x^2} \,  \big(G\xi\big)\cdot \big(G\xi\big)\  -\frac{1}{2}|\nabla\psi|_S^2\  -g\eta\\
&\ +\  \partial_x^{-1}\Big[\eta_x \mathcal{R}_x(x,z)\Big|_S\ -\    \mathcal{R}_z(x,z)\Big|_S \ -\  \omega_S(G\xi)\Big]\hspace{1.2cm} \textrm{on} \,\, S_{\eta},
\end{split}
\end{align}
 where
$\psi$ is given by \eqref{50}, the Dirichlet-Neumann operator has the expressions \eqref{G1}  and the function $\mathcal{R}(x,z)$ is given by
\begin{align}\label{R(x,z)}
\mathcal{R}(x,z)
:=&\ \int_{\Omega_\eta} \Big[ (G\xi)\, \Gamma_\eta(\mathbf{x},\mathbf{x}_0)\ \omega(\mathbf{x_0})\ +  \Gamma(\mathbf{x},\mathbf{x}_0)\  ( \psi_{x_0} \omega_{z_0}\ -\ \psi_{z_0}\omega_{x_0} ) \Big]d\mathbf{x}_0\nonumber\\
&\ +\int_{-\infty}^\infty \Gamma(\mathbf{x},x_0,\eta(x_0))\  \omega(x_0,\eta(x_0))\  G\xi (x_0) \  dx_0.
\end{align}
The expression \eqref{R(x,z)}  of the function $\mathcal{R}(x,z)$,  was obtained by taking into account \eqref{r}, the fact that the Green function $\Gamma(\mathbf{x},\mathbf{x}_0)$  depends explicitly on the free surface $\eta$ and the equations \eqref{evol-omega}, \eqref{KBC1} and \eqref{G} (for more details, see \cite{DIK-RI}).

\section{Point vortex motion coupled with the free surface motion}

We suppose
that the vorticity in the fluid domain  is concentrated
in a point
vortex submerged beneath the free surface, that is,
\begin{align}\label{omega*}
\omega\ =\ \omega^*\, \delta(\mathbf{x}-\mathbf{q})\ =\ \omega^*\delta(x-q_1)\delta(z-q_2),
\end{align}
where $\omega^*$ is a constant and $\mathbf{q}=(q_1(t),q_2(t))\in \Omega_\eta$ the position of the point vortex.
Then,
\begin{align}
\omega_t\ =\ \omega^*\Big(
-\delta'(x-q_1)\delta(z-q_2)\dot{q_1}\ -\  \delta(x-q_1)\delta'(z-q_2)\dot{q_2}
\Big)
\end{align}
and the first equation in
 \eqref{eqs1} becomes
\begin{align}\label{78}
\begin{split}
\dot{q_1}\ =&\ \psi_z\Big|_{\mathbf{x}=\mathbf{q}}\\
 \dot{q_2}\ =& -\psi_x\Big|_{\mathbf{x}=\mathbf{q}}.
\end{split}
\end{align}
Also from \eqref{omega*}, the expression of  the function $\psi$ in \eqref{50} reduces to
\begin{align}\label{30}
\psi(x,z)\ =&\  \omega^*  \Gamma(\mathbf{x},\mathbf{q})+i\   \sinh\big((z+h)D\big) \, \left[:\cosh\big((\eta+h)D\big):\right]^{-1}\xi,
\end{align}
and from \eqref{30}
\begin{align}\label{32}
\begin{split}
\psi_x\ =&\  \omega^*  \Gamma_x(\mathbf{x},\mathbf{q})
-   \sinh\big((z+h)D\big) D \left[:\cosh\big((\eta+h)D\big):\right]^{-1}\xi\\
\psi_z\ =&\ \omega^*  \Gamma_z(\mathbf{x},\mathbf{q})+i\   \cosh\big((z+h)D\big)D\left[:\cosh\big((\eta+h)D\big):\right]^{-1}\xi .
\end{split}\end{align}
 The Green function  diverges as a logarithm for  $\mathbf{x}\rightarrow\mathbf{q}$ and decays  to zero away
from the singularity. 
In the expressions of $\psi_x\Big|_{\mathbf{x}=\mathbf{q}}$ and  $\psi_z\Big|_{\mathbf{x}=\mathbf{q}}$,  we exclude the singular terms, we accept  as in \cite{M&Pulvirenti, Meleshko}, that there is no self-interaction, that is, the single vortex does not move under the action of its own
field. Other possibilities for regularization are explored in \cite{Holm}.\\
Another consequences of the vorticity in the   form \eqref{omega*} and of  the fact that we do not allow a point vortex  on the free surface are
the reduced expression for the function $\mathcal{R}(x,z)$ in \eqref{R(x,z)} (see \cite{DIK-RI})
\begin{align}\label{R(x,q)}
\mathcal{R}(x,z)=  \omega^*\Gamma_\eta(\mathbf{x}, \mathbf{q})\ G\xi+ \omega^*\Big(\Gamma_{q_1}(\mathbf{x},\mathbf{q})\dot{q}_1  + \Gamma_{q_2}(\mathbf{x},\mathbf{q})\dot{q}_2\Big),
\end{align}
and  $\omega_S=0$.
Therefore,
the system \eqref{eqs1} for the surface variables and the point vortex becomes (see \cite{DIK-RI})
\begin{align}\label{eqs_point_v}
 \begin{split}
\dot{q_1}\
=&\  i  \Big[ \cosh\big((z+h)D\big)D\left[:\cosh\big((\eta+h)D\big):\right]^{-1}\xi\Big]_{\mathbf{x}=\mathbf{q}}\\
 \dot{q_2}\
 =&\,\,\,\,  \Big[\sinh\big((z+h)D\big) D\left[:\cosh\big((\eta+h)D\big):\right]^{-1}\xi\Big]_{\mathbf{x}=\mathbf{q}}\\
r_t\ =&\  \mathcal{R}(x,z)\\
\eta_t\ =& \ G\xi \\
\xi_t\ =&\   \frac{\eta_x\,  \xi_x}{1+\eta_x^2}\,  G\xi  +\  \frac{1}{1+\eta_x^2} \,  \big(G\xi\big)\cdot \big(G\xi\big)\  -\frac{1}{2}|\nabla\psi|_S^2\  -g\eta\\
&\ +\  \partial_x^{-1}\Big[\eta_x \mathcal{R}_x(x,z)\Big|_S\ -\    \mathcal{R}_z(x,z)\Big|_S \Big]
\end{split}
\end{align}
where $\mathcal{R}(x,z)$ given by \eqref{R(x,q)} and
\begin{align}\label{83}
\hspace{-0.5cm}\begin{split}
\psi_x\Big|_S\ =&\   \omega^*  \Gamma_x(\mathbf{x},\mathbf{q})\Big|_S
-   :\sinh\big((\eta+h)D\big): D \left[:\cosh\big((\eta+h)D\big):\right]^{-1}\xi\\
\psi_z\Big|_S\ =&\ \omega^*  \Gamma_z(\mathbf{x},\mathbf{q})\Big|_S
+i: \cosh\big((\eta+h)D\big):D\left[:\cosh\big((\eta+h)D\big):\right]^{-1}\xi
\end{split}
\end{align}

Let  $\epsilon$ be  the ``small amplitude" parameter, which measures the ratio of the wave amplitude to the undisturbed fluid depth $h$, and $\delta= h/ \lambda$ - the shallowness parameter,  which measures the ratio of the undisturbed fluid depth $h$ to the wavelength $\lambda.$ % the magnitude of these parameters can be used to identify different general types of wave problems.
Now we consider small-amplitude long waves such that $\epsilon\ll 1$ and $\delta\ll 1$. The operator $D$ is of the order of
$\delta$, that is, $ h D \in \mathcal{O}(\delta)$, and the free surface elevation $\eta$ is of order $\epsilon,$ that is, $(\eta/ h) \in\mathcal{O}(\epsilon)$. Since $h$ is a constant, we will also use the notations $D \sim  \mathcal{O}(\delta),$   $\eta  \sim \mathcal{O}(\epsilon).$

We expand in Taylor series the operators on the right side of the equations
\eqref{eqs_point_v} and in  \eqref{83} and, by ignoring terms of orders  $\mathcal{O}(\epsilon^2, \epsilon\delta^2,\delta^4)$,
we get the following approximations (see \cite{DIK-RI})
\begin{align}
 \cosh &  \big((z+h)D\big) D\left[: \cosh\big((\eta+h)D\big):\right]^{-1}
\approx \  D,   \label{34}\\
 \sinh & \big((z+h)D\big) D\left[: \cosh\big((\eta+h)D\big):\right]^{-1} \approx\ (z+h) D^2 \label{35}
\end{align}
\begin{align}
G \xi  \ \approx \  D h D\xi\ -\ \frac{1}{3}Dh^3D^3 \xi\ +\ D\eta D \xi\ .
\end{align}
For the Taylor series expansion of the  Dirichlet-Neumann operator see also  \cite{Craig, Craig&Groves, CS}.
Therefore,  the first  two equations in \eqref{eqs_point_v} become
\begin{align}\label{99}
 \begin{split}
\dot{q_1}\
=&\   \xi_x (q_1) \\
 \dot{q_2}\
 =&-  (q_2+h)\xi_{xx}(q_1),
\end{split}
\end{align}
the  fourth equation in \eqref{eqs_point_v} takes the form
\begin{align}
\eta_t\ +\ h\, \xi_{xx}\ +\ \frac{1}{3}h^3 \xi_{xxxx}\ + (\eta\xi_x)_x\ =\ 0,
\end{align}
and for \eqref{83}  we find
\begin{align}\label{100}
\begin{split}
\psi_x\Big|_S\ =&\   \omega^*  \Gamma_x(\mathbf{x},\mathbf{q})\Big|_S
+  (\eta+h)\xi_{xx}(x)\\
\psi_z\Big|_S\ =&\ \omega^*  \Gamma_z(\mathbf{x},\mathbf{q})\Big|_S
+\xi_x(x).
\end{split}
\end{align}
In  small-amplitude long-wavelength regimes, the appropriate Green function for our point-vortex problem is the Green function on the infinite strip
$\{-\infty<x<\infty,\,   -h\ <\ z\ <0\}$. We use further  in the calculations the following Green function:
\begin{align}\label{Gamma-M-q}
\mathit{\Gamma}(\mathbf{x},\mathbf{q})=\frac{1}{4\pi}\ln\frac{\cosh\frac{\pi}{h}(x-q_1)-\cos\frac{\pi}{h}(z-q_2)}{
\cosh\frac{\pi}{h}(x-q_1)-\cos\frac{\pi}{h}(z+q_2)},
\end{align}
the formula obtained in  \cite{Melnikov} by the method of conformal mapping.
We see that far away from the point vortex, that is, $| \mathbf{x}-\mathbf{q} | \rightarrow \infty$, we have $\mathit{\Gamma}(\mathbf{x},\mathbf{q}) \rightarrow 0$,  and close to the vortex, that is,  $\mathbf{x}\rightarrow\mathbf{q}$, the Green function diverges as a logarithm.
 By accepting, as in \cite{M&Pulvirenti, Meleshko}, that there is no self-interaction, the above formula and the derivatives
\begin{align}
\begin{split}
 \mathit{\Gamma}_x(\mathbf{x},\mathbf{q})=\frac{1}{2h}\frac{\sin\left(\frac{\pi}{h}z\right) \sin\left(\frac{\pi}{h}q_2\right)\sinh\frac{\pi}{h}(x-q_1)}{
\big[\cosh\frac{\pi}{h}(x-q_1)-\cos\frac{\pi}{h}(z-q_2)\big]\big[\cosh\frac{\pi}{h}(x-q_1)-\cos\frac{\pi}{h}(z+q_2)\big]}\\
 \mathit{\Gamma}_z(\mathbf{x},\mathbf{q})=\frac{1}{2h}\frac{
\sin\left(\frac{\pi}{h}q_2\right)
\big[\cos\left(\frac{\pi}{h}q_2\right)-\cosh\frac{\pi}{h}(x-q_1) \cos\left(\frac{\pi}{h}z\right)\big]
}{
\big[\cosh\frac{\pi}{h}(x-q_1)-\cos\frac{\pi}{h}(z-q_2)\big]\big[\cosh\frac{\pi}{h}(x-q_1)-\cos\frac{\pi}{h}(z+q_2)\big]}\label{Gamma_x&z}
\end{split}
\end{align}
 apply everywhere   $\mathbf{x}\neq \mathbf{q}$. Therefore,
since $\eta  \sim \mathcal{O}(\epsilon)$,   by writing the Taylor series expansions of the functions  $ \mathit{\Gamma}_x(\mathbf{x},\mathbf{q})\Big|_S = \mathit{\Gamma}_x(x,\eta,q_1,q_2)$ and $ \mathit{\Gamma}_z(\mathbf{x},\mathbf{q})\Big|_S= \mathit{\Gamma}_z(x,\eta,q_1,q_2) $, we obtain
\begin{align}\label{113}
\begin{split}
 \mathit{\Gamma}_x(\mathbf{x},\mathbf{q})\Big|_S & = \mathcal{O}(\epsilon)\\
 \mathit{\Gamma}_z(\mathbf{x},\mathbf{q})\Big|_S &= - \frac{\sin\frac{\pi q_2}{h}}{
2h \left(\cosh \frac{\pi(x-q_1)}{h}-\cos \frac{\pi q_2}{h} \right)} +\mathcal{O}(\epsilon).
\end{split}
\end{align}
For the Boussinesq and KdV propagation regimes, we take $\epsilon \sim \delta^2$. The vortex strength corresponding to these regimes, depends on the vortex depth: if the point vortex is close to the free surface, then  $\omega^* \sim \mathcal{O}(\epsilon)$ and if the point vortex is near to the bottom, then
$\omega^* \sim \mathcal{O}(\delta^{\frac{3}{2}})$ (see \cite{DIK-RI}). \\
With \eqref{100} and \eqref{113} in view, we proceed with  the following approximations
\begin{align}
|\nabla\psi|^2_S\ \approx\ \Big[-\frac{\omega^*}{2h}\frac{\sin\frac{\pi q_2}{h}}{
 \left(\cosh \frac{\pi(x-q_1)}{h}-\cos \frac{\pi q_2}{h} \right)}+\xi_x(x)\Big]^2
\end{align}
and
\begin{align}\label{124}
\begin{split}
 -  &\eta_x \mathcal{R}_x(x,z)\Big|_S +    \mathcal{R}_z(x,z)\Big|_S \approx   - \frac{\omega^*\pi}{2h^2}\frac{\sin\frac{\pi q_2}{h}\sinh\frac{\pi(x-q_1)}{h}}{\left(\cosh\frac{\pi(x-q_1)}{h}-\cos\frac{\pi q_2}{h}\right)^2} \, \xi_x(q_1)
\\
&\hspace{4cm} - (q_2+h)\frac{\omega^*\pi}{2h^2}\frac{1-\cos\frac{\pi q_2}{h}\cosh\frac{\pi(x-q_1)}{h}}{\left(\cosh\frac{\pi(x-q_1)}{h}-\cos\frac{\pi q_2}{h}\right)^2}\xi_{xx}(q_1) \\
%&=- \frac{\omega^*\pi\left[\sin\frac{\pi q_2}{h}\sinh\frac{\pi(x-q_1)}{h}\mathfrak{u}(q_1) +  (q_2\! +\! h)(1-\cos\frac{\pi q_2}{h}\cosh\frac{\pi(x-q_1)}{h})\mathfrak{u}_{x}(q_1)\right]}{2h^2\left(\cosh\frac{\pi(x-q_1)}{h}-\cos\frac{\pi q_2}{h}\right)^2}
\end{split}
\end{align}
(see \cite{DIK-RI} for more details).
With  the notation
\begin{align}\label{109}
\mathfrak{u}:=\xi_x \sim \mathcal{O}(\epsilon),
\end{align}
which yields
\begin{align}\label{110}
\mathfrak{u}_x=\xi_{xx} \sim \mathcal{O}(\epsilon\delta),
\end{align}
we obtain in the regime $\epsilon \sim \delta^2$, the following system for the point vortex motion coupled with  the free surface motion
\begin{align}\label{system-final}
 \begin{split}
& \dot{q_1}\
=  \mathfrak{u} (q_1(t),t)\\
& \dot{q_2}\
 = - (q_2+h)\mathfrak{u}_{x}(q_1)\\
& \eta_t\ +\ h\mathfrak{u}_{x}\ +\   \frac{1}{3}h^3\mathfrak{u}_{xxx}\ +\  (\eta\mathfrak{u})_x=0\\
& \mathfrak{u}_t\ +\ g\eta_x\ +\ \frak{N}([\frak{u}],\mathbf{q})=0.
\end{split}
\end{align}
where we have denoted  by
\begin{align}\label{N}
\hspace{-0.7cm}\begin{split}
\frak{N}([\frak{u}],\mathbf{q}):=& \left\{\frac{1}{2}\left[\mathfrak{u}-\omega^*\frac{\sin\frac{\pi q_2}{h}}{
2h \left(\cosh \frac{\pi(x-q_1)}{h}-\cos \frac{\pi q_2}{h} \right)}\right]^2 \right \}_x\\
&\hspace{0cm}- \frac{\omega^*\pi}{2h^2}\left\{\frac{\sin\frac{\pi q_2}{h}\sinh\frac{\pi(x-q_1)}{h}}{\left(\cosh\frac{\pi(x-q_1)}{h}-\cos\frac{\pi q_2}{h}\right)^2} \, \mathfrak{u}(q_1)\right.\\
&\hspace{2cm}\left. + (q_2+h)\frac{1-\cos\frac{\pi q_2}{h}\cosh\frac{\pi(x-q_1)}{h}}{\left(\cosh\frac{\pi(x-q_1)}{h}-\cos\frac{\pi q_2}{h}\right)^2}\mathfrak{u}_{x}(q_1) \right\}
\end{split}
\end{align}
At the leading order in $\epsilon $, the last two  equations  in \eqref{system-final} are:
\begin{equation}
    \begin{split}
       \eta_t &+ h \mathfrak{u}_x=0 \\
       \mathfrak{u}_t &+ g \eta_x=0. 
    \end{split}
\end{equation}
Since $\eta$ and  $ \mathfrak{u}$  satisfy  the wave equation, i.e., $\eta_{tt}-gh \eta_{xx}=0$ and  $\mathfrak{u}_{tt}-gh \mathfrak{u}_{xx}=0$, respectively,  there are only travelling waves with speed
\begin{align}
c^2=gh.
\end{align}
Without loss of generality, we study the right-running waves assuming $c=\sqrt{gh}.$    Thus, for  $\eta=H(x-ct)$ and $\mathfrak{u}=U(x-ct)$, we get 
\begin{equation}
    \begin{split}
       -cH'\ +\ hU'\ &=\ 0 \\
       -cU'\ +\ gH'\ &=\ 0. 
    \end{split}
\end{equation}
 The compatibility of these equations gives again $c^2=gh$ and,  for waves which decay fast to zero at $\pm \infty,$  we get
 \begin{align}
H=\frac{h}{c}U.
\end{align}
 Therefore, at the leading order in $\epsilon$,  we have  
\begin{align}\label{speed}
\eta=\frac{h}{c} \mathfrak{u}.
\end{align}
 The KdV type equation appears at the next order approximation in  $\epsilon$. 
For $\eta,\, \mathfrak{u},\, \omega^*\sim \mathcal{O}(\epsilon)$,  $\eta_t, \, \mathfrak{u}_x\sim \mathcal{O}(\epsilon\delta)$, $\mathfrak{u}_{xxx},\, (\eta\mathfrak{u})_x \sim \mathcal{O}(\epsilon^2\delta)$,  in the last two equations in \eqref{system-final} let us explicitly introduce the scale parameters:
\begin{align}\label{CV1}
 \begin{split}
& \eta_t + h\mathfrak{u}_{x} +  \epsilon \frac{1}{3}h^3\mathfrak{u}_{xxx}\ + \epsilon (\eta\mathfrak{u})_x=0\\
& \mathfrak{u}_t +g\eta_x\
+ \epsilon \, \frak{N}([\frak{u}],\mathbf{q})=0.
\end{split}
\end{align}
To obtain the KdV type equation, we can follow Section 6 in  \cite{DIK-RI} 
or an alternative way presented below.\\
  Taking into account the expressions of the terms entering in the system \eqref{CV1},  let us  express $\eta$ 
  via $\mathfrak{u}$ by a  transformation of the following form:
\begin{equation}\label{63}
    \eta=\frac{h}{c}\mathfrak{u}+ \epsilon A\, \mathfrak{u}_{xx}+\epsilon \frak{B} ([\mathfrak{u}], \mathbf{q}),
\end{equation} for an yet unknown constant $A$ and functional $\frak{B}$ depending on $\mathfrak{u}$ and $\mathbf{q}$. 
%Therefore,
%\begin{align}\label{64}
%\begin{split}
%\eta_t&=\frac{h}{c}\mathfrak{u}_t+ \epsilon A\, %\mathfrak{u}_{xxt}+\epsilon \frak{B}_t,\\
%\eta_x&=\frac{h}{c}\mathfrak{u}_x+ \epsilon A\,  %\mathfrak{u}_{xxx}+\epsilon \frak{B}_x.
%\end{split}
%\end{align}
We  substitute \eqref{63} in the equations \eqref{CV1}, we take \eqref{speed} into account, and,  by neglecting the terms of order $\epsilon^2$, 
we get
\begin{align}\label{65}
\begin{split}
\mathfrak{u}_t& + c\mathfrak{u}_x+ \epsilon\frac{c A}{h} \mathfrak{u}_{xxt}+  \epsilon \frac{ch^2}{3}\mathfrak{u}_{xxx}+\epsilon\frac{c}{h} \frak{B}_t  + \epsilon(\mathfrak{u}^2)_x=0\\
\mathfrak{u}_t& + c\mathfrak{u}_{x} + \epsilon \frac{c^2A}{h}    \mathfrak{u}_{xxx} + \epsilon \Big[\frac{c^2}{h}  \frak{B}_x+\frak{N}([\frak{u}],\mathbf{q})\Big] =0.
\end{split}
\end{align}
The equations \eqref{65}  suggest  that there exists a slow time variable, such that 
\begin{equation}
\partial_t + c \partial_x = \epsilon \partial_T,\quad \partial_x=\partial_X,
\end{equation}
 that is, a Gallilean change of variables 
\begin{align}\label{XT}
X = x-ct, \quad T=\epsilon t.
\end{align}
Therefore, at the leading order, the equations \eqref{65}  become
\begin{align}
\begin{split}
 \mathfrak{u}_T&+\left(-\frac{c^2 A}{h}+ \frac{ch^2}{3}\right) \mathfrak{u}_{XXX}-\frac{c^2}{h}\frak{B}_X+ (\mathfrak{u}^2)_X=0\\
\mathfrak{u}_T&  + \frac{c^2A}{h}  \,  \mathfrak{u}_{XXX} + \frac{c^2}{h}  \frak{B}_X+\frak{N}=0.
\end{split}
\end{align} 
These equations  should be identical, which means that the coefficients in front of $\mathfrak{u}_{xxx}$ should coincide. This yields
\begin{align}
\begin{split}
A&= \frac{h^3}{6c}\\
\frak{B}([\frak{u}],\mathbf{q})&=\frac{h}{2c^2}\Big(\frak{u}^2-\partial_x^{-1}\frak{N}\Big),
\end{split}
\end{align}
since $\partial_X=\partial_x$. Finally, \eqref{63}  and \eqref{65}  acquire the form (without the scaling)
 \begin{equation} \label{eta_d}
    \eta=\frac{h}{c}\mathfrak{u}+ \frac{h^3}{6c} \mathfrak{u}_{xx}+ 
\frac{h}{2c^2}\Big(\frak{u}^2-\partial_x^{-1}\frak{N}\Big),
\end{equation}    
    \begin{equation}
  \mathfrak{u}_t\ + c\mathfrak{u}_{x}\ + \frac{c h^2}{6} \mathfrak{u}_{xxx} + \frac{1}{2}  \Big[(  \mathfrak{u}^2)_x +\frak{ N}\Big]=0.
    \end{equation}
Summing up, we obtain the following perturbed KdV equation for $\mathfrak{u}(x,t)$ coupled with the dynamic equations for the point vortex $\left(q_1(t),q_2(t)\right)$:
\begin{align}\label{KdV system q}
 \begin{split}
& \dot{q_1}\
=  \mathfrak{u} (q_1(t),t)\\
&
 \dot{q_2}\
 =-  (q_2+h)\mathfrak{u}_{x}(q_1(t),t)\\
&    \mathfrak{u}_t  +c   \mathfrak{u}_x+ \frac{c h^2}{6}\mathfrak{u}_{xxx}\\
&\hspace{0.7cm}+
\left\{\frac{1}{4}\left[\mathfrak{u}-\omega^*\frac{\sin \frac{\pi q_2}{h}}{2h \left(\cosh \frac{\pi(x-q_1)}{h}-\cos \frac{\pi q_2}{h} \right)}\right]^2 +\frac{1}{2} \mathfrak{u}^2 \right\}_{x}\\
&\hspace{0.7cm}-\frac{\omega^*\pi}{4h^2}\left\{\frac{\sin\frac{\pi q_2}{h}\sinh\frac{\pi(x-q_1)}{h}}{\left(\cosh\frac{\pi(x-q_1)}{h}-\cos\frac{\pi q_2}{h}\right)^2} \, \mathfrak{u}(q_1)\right.\\
&\hspace{2cm}\left.+ (q_2+h)\frac{1-\cos\frac{\pi q_2}{h}\cosh\frac{\pi(x-q_1)}{h}}{\left(\cosh\frac{\pi(x-q_1)}{h}-\cos\frac{\pi q_2}{h}\right)^2}\mathfrak{u}_{x}(q_1) \right\}  =0.
\end{split}
\end{align}

The integration in \eqref{eta_d} can be performed explicitly and the result is the following expression for the surface elevation:
\begin{align} \label{eta final}
    \eta(x,t)&=\frac{h}{c}\mathfrak{u}(x,t)+  \frac{h^3}{6c} \mathfrak{u}_{xx}(x,t)+ 
\frac{h}{2c^2} \frak{u}^2 (x,t) \nonumber \\
&  -\frac{h}{4c^2} \left[\mathfrak{u}(x,t)-\omega^*\frac{\sin\frac{\pi q_2}{h}}{
2h \left(\cosh \frac{\pi(x-q_1)}{h}-\cos \frac{\pi q_2}{h} \right)}\right]^2  \nonumber \\
&\hspace{0cm} + \frac{\omega^* }{4c^2}    \left\{ - \mathfrak{u}(q_1(t),t)  \frac{\sin\frac{\pi q_2}{h}}{\cosh\frac{\pi(x-q_1)}{h}-\cos\frac{\pi q_2}{h} } \, \right.  \nonumber \\
&\hspace{2cm}\left. + (q_2(t)+h)\mathfrak{u}_{x}(q_1(t),t)\frac{ \sinh\frac{\pi(x-q_1)}{h}}{\left(\cosh\frac{\pi(x-q_1)}{h}-\cos\frac{\pi q_2}{h}\right)}\right\} .
\end{align}

\section{Analytical investigation of the vortex motion influenced by unperturbed KdV solitons}

Before investigating the coupled system \eqref{KdV system q} numerically, we analyse some features provided by the simplification that the KdV soliton solutions travel unperturbed above the point vortex, and the vortex moves in the velocity field created by the soliton. In essence, this is a ``regular fluid particle" approximation for the motion of the point vortex. In Section \ref{numerics}, it will be confirmed that such decoupling is justifiable. The particle dynamics in the KdV approximation has previously been studied in \cite{Bor}.   

In this approximation, $\omega^*=0$, $\mathfrak{u}(x,t)$ satisfies the standard KdV equation 
\begin{equation} \label{KdV 0}
      \mathfrak{u}_t  +c   \mathfrak{u}_x+ \frac{c h^2}{6}\mathfrak{u}_{xxx}+ \frac{3}{2}\mathfrak{u}\mathfrak{u}_x =0,
\end{equation}
and the surface elevation $\eta(x,t)$ is given by 
\begin{align} \label{eta 68}
    \eta(x,t)&=\frac{h}{c}\mathfrak{u}(x,t)+  \frac{h^3}{6c} \mathfrak{u}_{xx}(x,t)+ 
\frac{h}{4c^2} \frak{u}^2 (x,t). 
\end{align}
The vortex motion is governed by the dynamical system 
\begin{align}\label{KdV system q free}
 \begin{split}
& \dot{q_1}\
=  \mathfrak{u} (q_1(t),t)\\
&
 \dot{q_2}\
 =-  (q_2+h)\mathfrak{u}_{x}(q_1(t),t)
 \end{split}
 \end{align}
In the next order approximation, by keeping the next order terms from \eqref{34} and \eqref{35} in the expansions of the operators, the system \eqref{KdV system q free} becomes 
\begin{equation}\label{q system epsilon}
    \begin{split}
         & \dot{q_1}\
=  \mathfrak{u} (q_1(t),t) -  \frac{q_2(q_2+2h)}{2}  \mathfrak{u}_{xx} (q_1(t),t)\\
&
 \dot{q_2}\
 =-  (q_2+h)\mathfrak{u}_{x}(q_1(t),t) +  \frac{(q_2+h)(q_2^2+2hq_2-2h^2)}{6}  \mathfrak{u}_{xxx} (q_1(t),t) 
    \end{split}
\end{equation}
  The additional terms are of smaller order with respect to $\epsilon$ in comparison to the leading order terms. In Section \ref{4.1} and Section \ref{4.2} it will be shown that their presence has very little effect on the dynamics of the point vortex / fluid particles.

\subsection{Exact KdV  one-soliton approximation}\label{4.1}

 The exact one-soliton solution of the KdV equation \eqref{KdV 0} is given by the expression (see, for example, \cite{Cullen&Ivanov})
\begin{equation}\label{KdVsol}
\mathfrak{u}(x,t)=\frac{4 c h^2 K^2}{3 \cosh^2\left[K\left(x-x_0-(1+\frac{2}{3}h^2K^2)\, ct\right)\right]},
\end{equation} where $x_0$ and $K$, called the soliton parameters, are constants.   The constant $x_0$  determines an
initial  position of the soliton at $t = 0$. The above soliton propagates at a constant speed $(1+\frac{2}{3}h^2K^2) c.$ The leading order approximation for the wave speed is $c=\sqrt{gh}$ and the first correction is proportional to the $K^2h^2$ term.
The consistency with the KdV approximation requires that the choice of the soliton parameter $K$ is of the order of the wave number in the long-wave regime. In other words, $Kh$ should be a quantity of order $\delta,$  the shallowness parameter. In this case, the amplitude of the soliton is also proportional to $K^2h^2$ and therefore the soliton has a magnitude of order $\epsilon$ (the small-amplitude parameter) which agrees with our assumptions. \\
At the leading order,
\begin{align}\label{KdV eta leading order}
\eta(x,t)=\frac{h}{c}\mathfrak{u}(x,t)=\frac{4  h^3 K^2}{3 \cosh^2\left[K\left(x-x_0-(1+\frac{2}{3}h^2K^2)\, ct\right)\right]},
\end{align}
 the next order approximation includes the terms as in  \eqref{eta 68}, where $\mathfrak{u}(x,t)$  is now given by \eqref{KdVsol}; the higher-order terms are not written explicitly because their expressions are very elaborate. In order to see the differences between the two cases,  snapshots of the free surface \eqref{KdV eta leading order} are plotted (in black) for different values of $t$ in Fig. \ref{fig:eta}, as well as of the free surface \eqref{eta 68}  with the higher-order terms (in red).

 For the exact one-soliton solution \eqref{KdVsol}, 
the two equations \eqref{KdV system q free} which determine the vortex trajectory in the ``regular fluid particle" approximation, reduce to
\begin{align}\label{qKdVsoliton}
 \begin{split}
& \dot{q_1}= \frac{4 c h^2 K^2}{3 \cosh^2\left[K\left(q_1-x_0-(1+\frac{2}{3} h^2K^2)\, ct\right)\right]}\\
&
 \dot{q_2} =  (q_2+h)\frac{8 c h^2 K^3  \sinh \left[K\left(q_1-x_0-(1+\frac{2}{3} h^2K^2)\, ct\right)\right] }{3 \cosh^3\left[K\left(q_1-x_0-(1+\frac{2}{3}h^2K^2)\, ct\right)\right]}.
\end{split}
\end{align}
The next-order approximation terms are as in \eqref{q system epsilon}, with $\mathfrak{u}(x,t)$  given by \eqref{KdVsol}. Their expressions are straightforward, although very elaborate, and are used for the plots of Fig. \ref{fig:different_q2} -   Fig. \ref{fig:different_q1}.\\
For the system \eqref{qKdVsoliton}, we observe that the trajectories are bounded in a finite domain in space. The horizontal motion is within a finite interval and the extent of this interval can be obtained analytically. By
integrating the first equation in the system above, we obtain

{\footnotesize \begin{align}\label{cl}
\hspace{-0.3cm} q_1(t)+\frac{2h}{\sqrt{3-2h^2K^2}  } \textrm{arctan}\left[ \frac{2hK \tanh \left[K\left(q_1(t)-x_0-(1+\frac{2}{3}h^2K^2)\, ct\right)\right]    }{\sqrt{3-2h^2K^2}} \right]=\text{const.}
\end{align}}
Therefore,  by taking into account that $c\in\mathcal{O}(1)$, $h^2K^2\in \mathcal{O}(\epsilon)$, $\dot{q_1}\in \mathcal{O}(\epsilon)$, we go to the limits $t\to \pm \infty$ in \eqref{cl} and find the following,
\begin{align}\label{range}
q_1(\infty)-q_1(-\infty)= \frac{4h}{\sqrt{3-2h^2K^2}} \textrm{arctan}\left[\frac{2hK}{\sqrt{3-2h^2K^2}}\right].
\end{align}

%\textcolor{blue}{(I rewrote the text after (56))\\
Although the second equation in \eqref{qKdVsoliton} is much more difficult to integrate, we can also estimate the extent of vertical motion. Since at the leading order, the derivative with respect to $x$ is the  derivative with respect to $t$ multiplied by $-\frac{1}{c}$ (since $\mathfrak{u}_x =-\frac{1}{c} \mathfrak{u}_t+...$), the second equation in \eqref{KdV system q} can be written, in the leading order approximation, as  
\begin{align}\label{60}
\dot{q_2}\
 =\frac{1}{c} (q_2+h)\mathfrak{u}_{t}(q_1(t),t).
    \end{align}
    At the initial time $t=0$, the soliton is exactly above the vortex, that is, 
    \begin{align}\label{61}
    x_0=q_1(0),
    \end{align}
    and in view of \eqref{KdVsol} 
    \begin{equation}\label{62}
\mathfrak{u}(q_1(0),0)=\frac{4 c h^2 K^2}{3}.
\end{equation}
By integrating the equation \eqref{60} from $0$ to $\infty$ and from $-\infty$ to $0$, taking into account \eqref{61}-\eqref{62} and the fact that, at the locations $q_1(\pm\infty)$ we have $\mathfrak{u}(q_1(\pm \infty),\pm \infty) = 0$, we get 
\begin{align}
    q_2(\pm \infty) + h\ =\ [q_2(0)+h]\exp\left(-\frac{4h^2K^2}{3}\right).
\end{align}
Therefore, the vertical motion is within a finite interval, the extent of this interval is approximately given by
\begin{equation}\label{vert_shift}
     q_{2}(0)- q_{2}(\pm \infty)=( q_{2}(0)+h )\left[1-\exp(-4K^2 h^2 /3)\right].
\end{equation}
%}
The plots in Fig. \ref{fig:different_q2} -- Fig. \ref{fig:different_q1} include the graph of $q_1$ versus $t$, the graph of $q_2$ versus $t$, and the graph of $q_2$ versus $q_1.$  They are plotted for the following parameter values:
 $x_0=0$, $h=10$ m, $g=9.81$ m/s$^2$ and  $K=0.05$ m$^{-1}.$  The figures in the left panels of  Fig. \ref{fig:different_q2}, labeled with sub-indices $a$ are the plots for the system \eqref{qKdVsoliton}, the the figures in the right panels, labeled with sub-indices $b$ are the plots for the system \eqref{q system epsilon} with the higher-order terms,  $\mathfrak{u}(x,t)$  given by \eqref{KdVsol}. In Fig. \ref{fig:different_q2}, we take the same initial condition  $q_1(0)=0$ and  the initial condition for $q_2$   takes different values:  $q_2(0)=-1$ (blue), $q_2(0)=-5$ (orange), $q_2(0)=-8$ (green).
 In Fig. \ref{fig:different_q1}, we take the same initial condition  $q_2(0)=-1$ and  the initial condition for $q_1$   takes different values:  $q_1(0)=5$ (blue), $q_1(0)=0$ (orange), $q_1(0)=-8$ (green). The trajectories are extended for both positive and negative $t.$
 We can see in  Fig. \ref{fig:different_q2}, {panel ($iii_a$) and Fig. \ref{fig:different_q1}, panel ($iii_a$) that  the point vortex trajectories are bounded and the extent of the interval for the horizontal motion is in agreement with  the value in \eqref{range}, that is, $ \frac{40}{\sqrt{2.5}} \textrm{arctan}\left[\frac{1}{\sqrt{2.5}}\right]\approx 14.3 $ m. The vertical shift according to the approximation \eqref{vert_shift} for the trajectories on  Fig. \ref{fig:different_q2}, panel ($iii_a$) is 2.6 m (blue trajectory), 1.4 m (orange trajectory) and 0.57 m (green trajectory), which is in a very good agreement with the numerical output. By comparing the panels with label a with those with label b, we notice that the differences are almost imperceptible. When higher order terms are taken into account, the plots are in the panels with label b; the trajectories are very slightly flattened, but the shape of the trajectories and their boundedness are preserved.\\
In Fig. \ref{fig:different_q2 and eta} we also plot in red  the evolution in time of the free surface $\eta$ for $x=0$ on top of the evolution in time of $q_2$, for the same values as those used for the plot in Fig. \ref{fig:different_q2}. In panel (a), the free surface $\eta$ is given by \eqref{KdV eta leading order}, while in panel (b) the higher order terms as in \eqref{eta 68} are taken into account, with $\mathfrak{u}(x,t)$  given by \eqref{KdVsol}. 
 \begin{figure}[h!]
\begin{center}
\fbox{\includegraphics[totalheight=0.085\textheight]{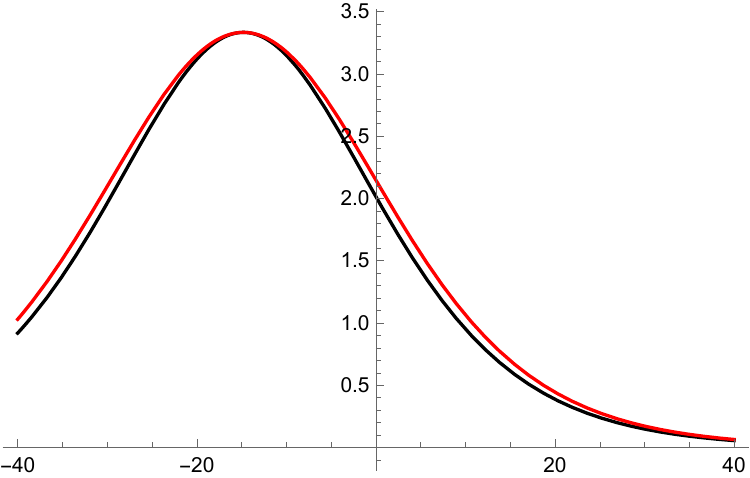}}
\fbox{\includegraphics[totalheight=0.085\textheight]{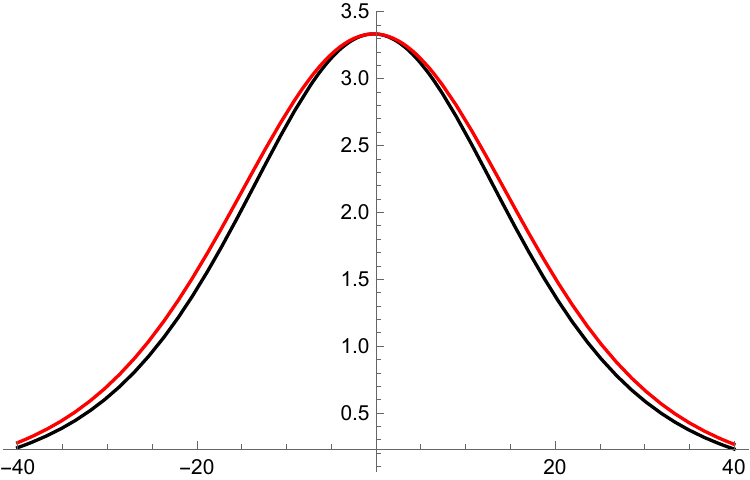}}
\fbox{\includegraphics[totalheight=0.085\textheight]{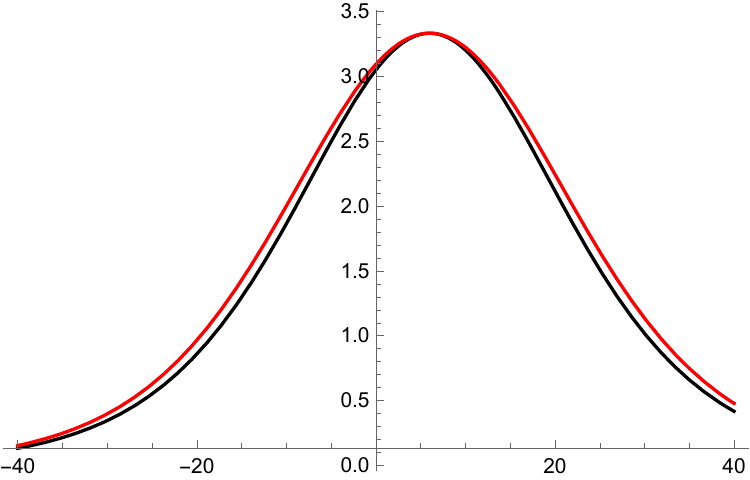}}
\fbox{\includegraphics[totalheight=0.085\textheight]{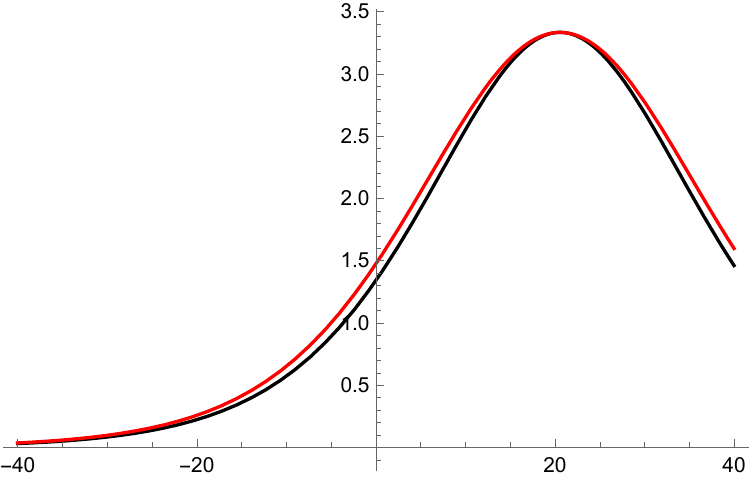}}
\caption{
Snapshots  of the free surface \eqref{KdV eta leading order} in black, as well as of the free surface \eqref{eta 68} with the appropriate higher-order terms for $\mathfrak{u}(x,t)$  given by \eqref{KdVsol}, in red.
The function $\eta(x,t)$ is plotted as a function of $x$ for different values of $t.$ The parameters are $x_0=0$,  $h=10$m,  $K=0.05$m$^{-1}$.}
\label{fig:eta}
\end{center}
\end{figure}

\begin{figure}[h!]
\begin{center}
(i$_a$)\fbox{\includegraphics[totalheight=0.16\textheight]{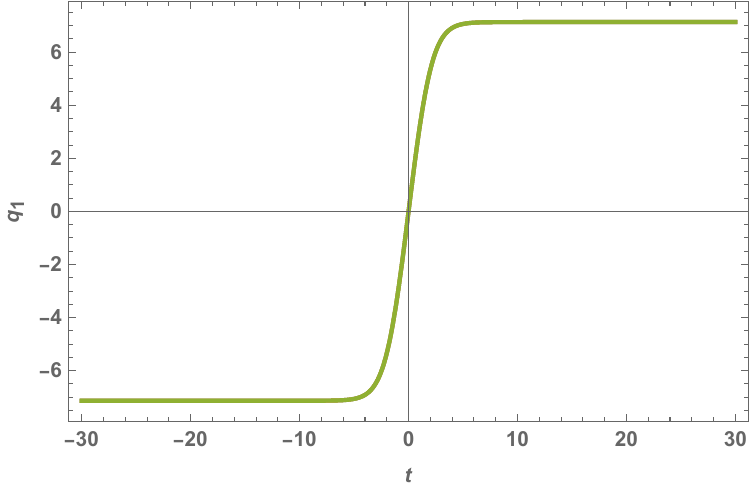}}
(i$_b$)\fbox{\includegraphics[totalheight=0.16\textheight]{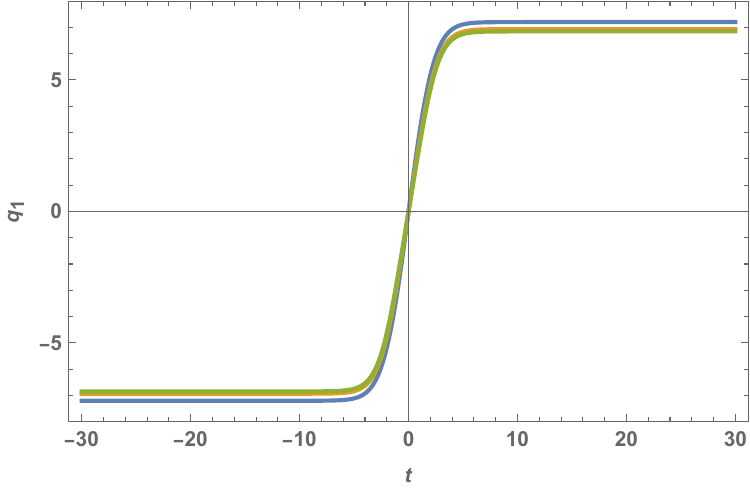}}
(ii$_a$)\fbox{\includegraphics[totalheight=0.16\textheight]{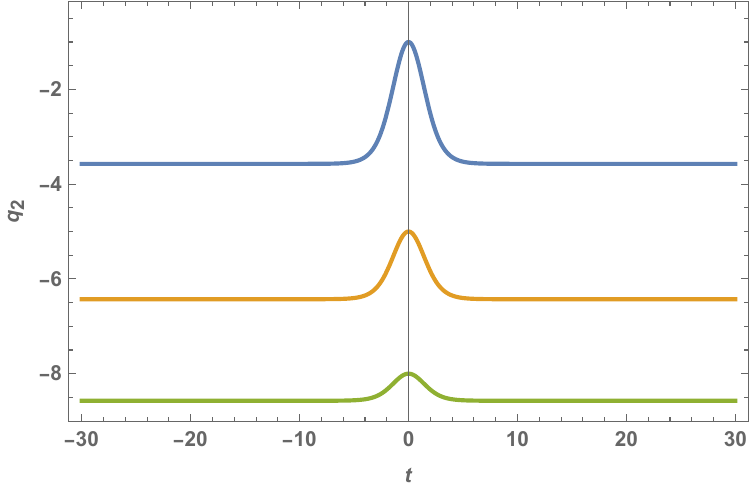}}
(ii$_b$)\fbox{\includegraphics[totalheight=0.16\textheight]{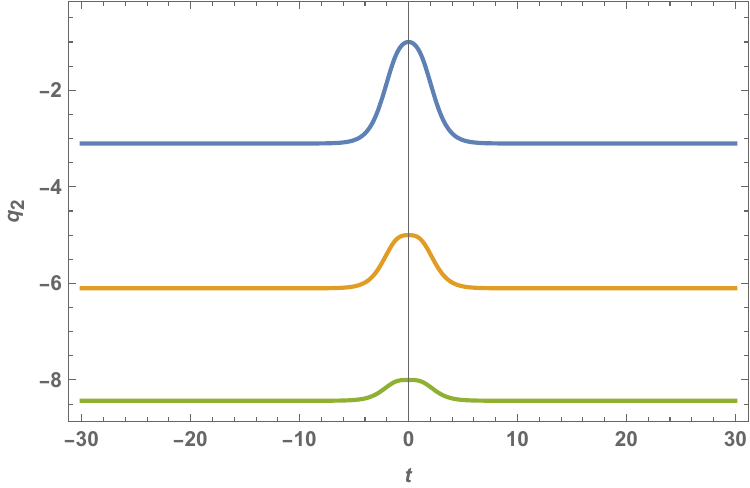}}
(iii$_a$)\fbox{\includegraphics[totalheight=0.14\textheight]{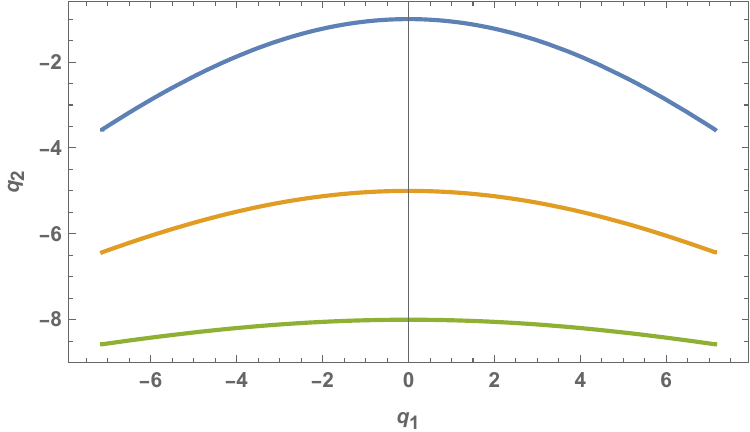}}
(iii$_b$)\fbox{\includegraphics[totalheight=0.14\textheight]{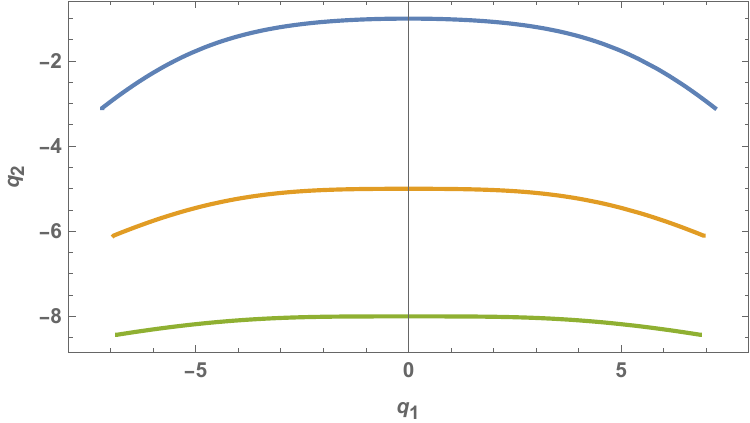}}
\caption{ (i) The evolution in time of $q_1$; (ii) The evolution in time of $q_2$;
(iii) Vortex trajectories at different depth. The panels labeled with a sub-index $a$ are plots for the system \eqref{qKdVsoliton}, the panels labeled with a sub-index $b$ are plots for the system \eqref{q system epsilon} with the higher-order terms,  $\mathfrak{u}(x,t)$  given by \eqref{KdVsol}. The parameter values are $x_0=0$,  $h=10$m,  $K=0.05$m$^{-1}$,  the initial condition $q_1(0)=0$  and the initial condition  $q_2$ takes different values: $q_2(0)=-1$ (blue), $q_2(0)=-5$ (orange), $q_2(0)=-8$ (green).}
\label{fig:different_q2}
\end{center}
\end{figure}
\begin{figure}[h!]
	\begin{center}
		($a$)\fbox{\includegraphics[totalheight=0.16\textheight]{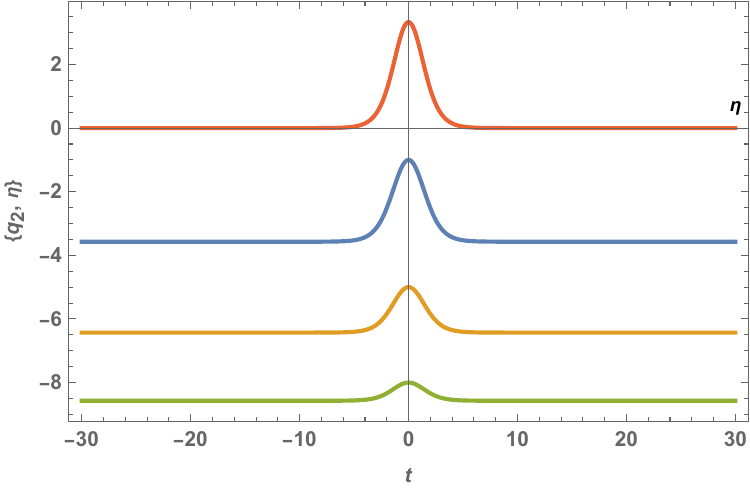}}
		($b$)\fbox{\includegraphics[totalheight=0.16\textheight]{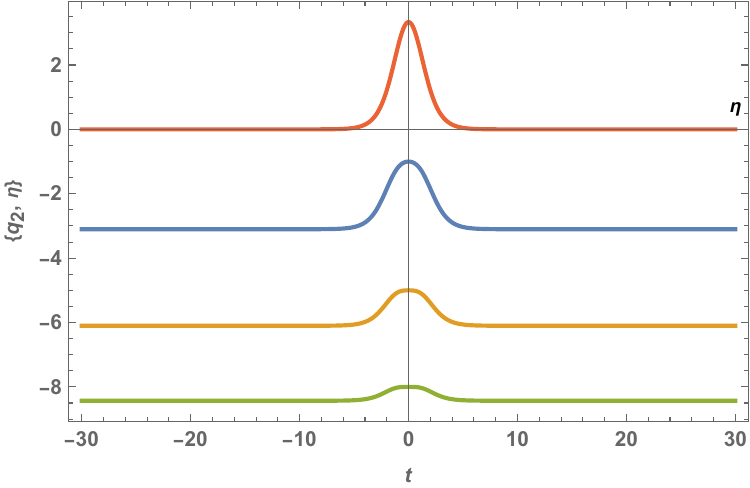}}
				\caption{The time evolution of the free surface $\eta$  for $x=0$ (red)  and of $q_2$, for the same parameter values as those used in Figure  \ref{fig:different_q2}.  (a) $\eta$ in the leading order, given by  \eqref{KdV eta leading order}; (b) $\eta$ with the higher-order terms in \eqref{eta 68} with $\mathfrak{u}(x,t)$  given by \eqref{KdVsol}.}
		\label{fig:different_q2 and eta}
	\end{center}
\end{figure}
\begin{figure}[h!]
\begin{center}
(i$_a$)\fbox{\includegraphics[totalheight=0.165\textheight]{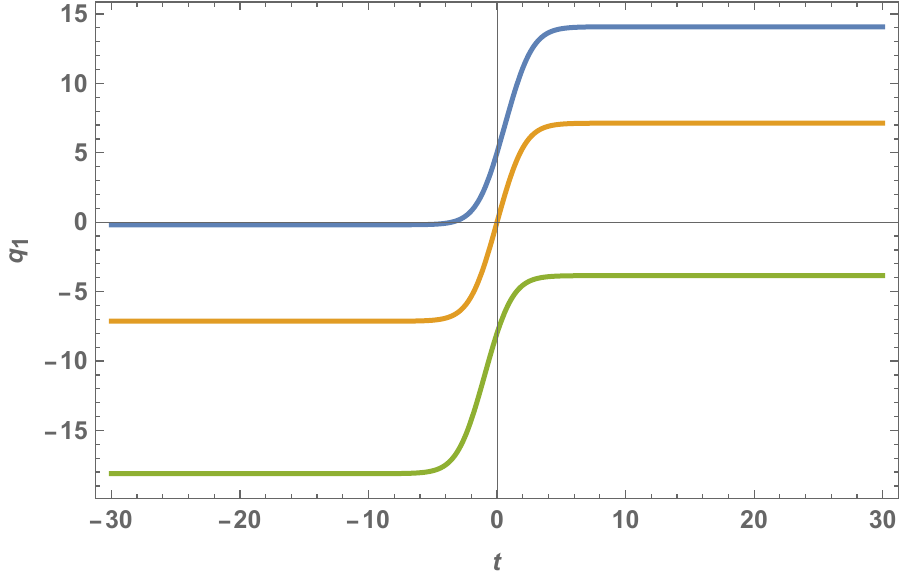}}(i$_b$)\fbox{\includegraphics[totalheight=0.165\textheight]{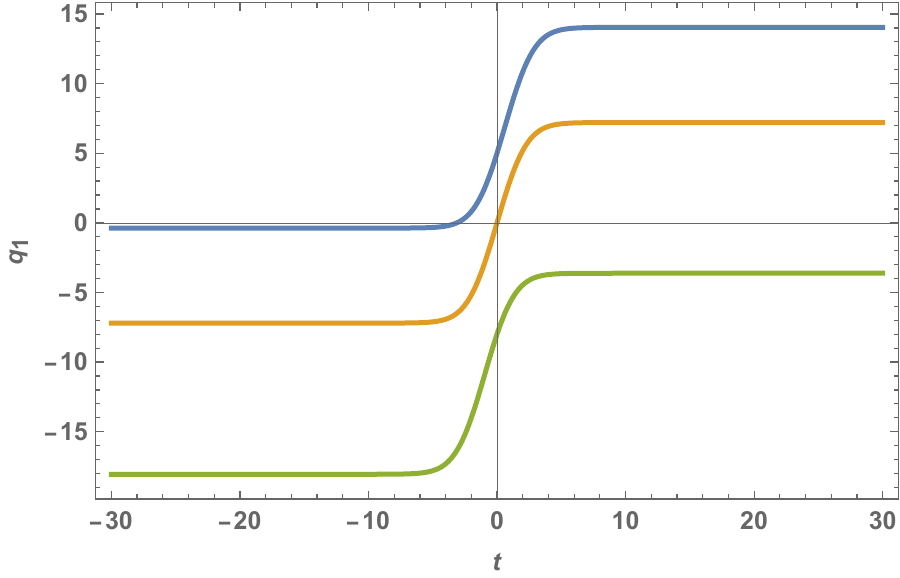}}
(ii$_a$)\fbox{\includegraphics[totalheight=0.16\textheight]{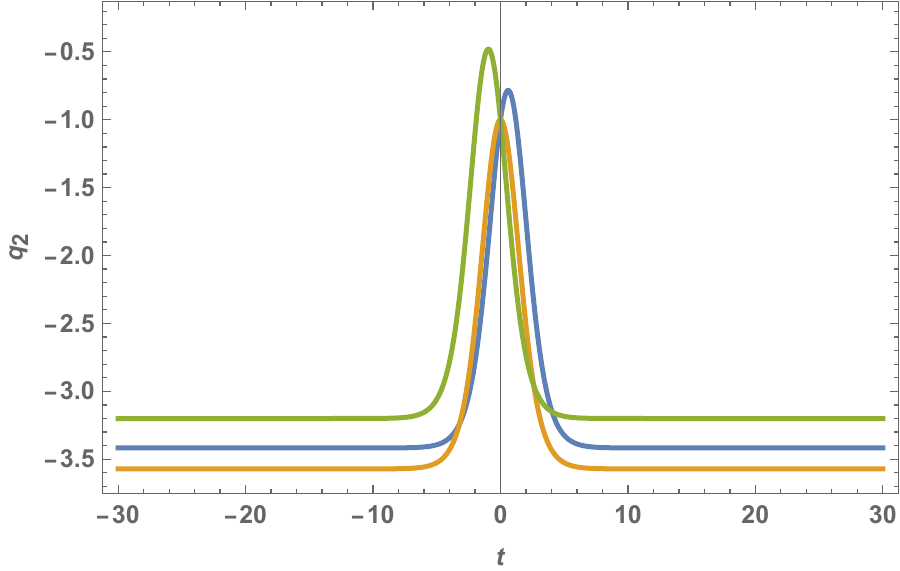}}
(ii$_b$)\fbox{\includegraphics[totalheight=0.16\textheight]{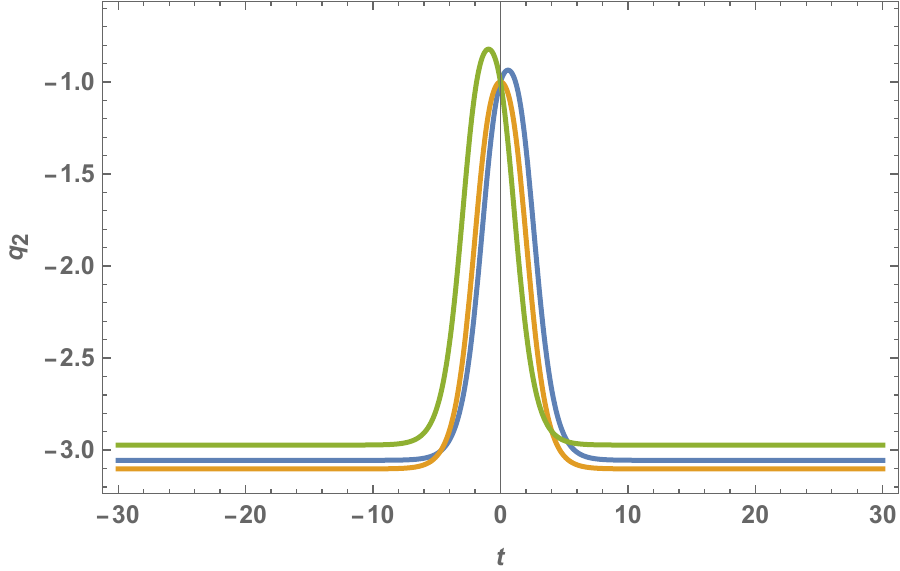}}
(iii$_a$)\fbox{\includegraphics[totalheight=0.085\textheight]{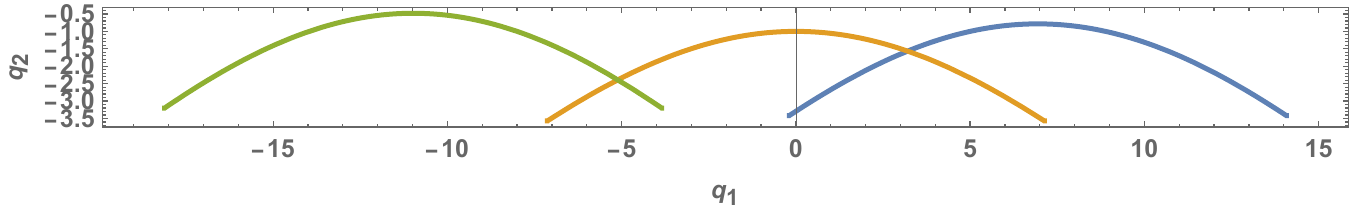}}
(iii$_b$)\fbox{\includegraphics[totalheight=0.07\textheight]{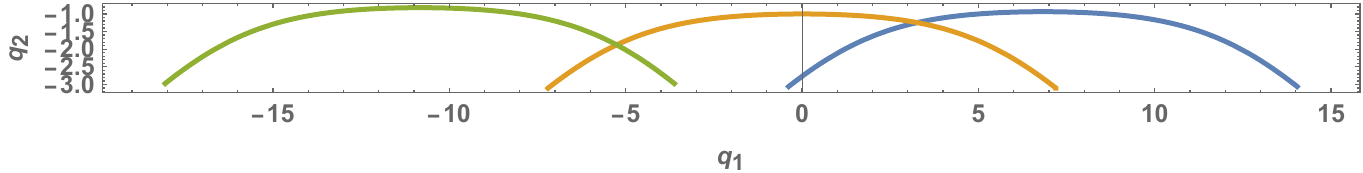}}
\caption{
(i) The time evolution of $q_1$; (ii) The time evolution of $q_2$;
(iii)  Vortex trajectories at the same depth. The panels labeled with a sub-index $a$ are plots for the system \eqref{qKdVsoliton}, the panels with the sub-index $b$  are plots for the system \eqref{q system epsilon} with the higher-order terms,  $\mathfrak{u}(x,t)$  is given by \eqref{KdVsol}. In all plots  $x_0=0$,  $h=10$m,  $K=0.05$m$^{-1}$,  the initial condition  $q_1$ takes different values: $q_1(0)=5$ (blue), $q_1(0)=0$ (orange), $q_1(0)=-8$ (green) and the initial condition  $q_2(0)=-1$.}
\label{fig:different_q1}
\end{center}
\end{figure}

\subsection{Exact KdV two-soliton approximation}\label{4.2}

%\textcolor{blue}{Now we analyse the system \eqref{KdV system q} with $\omega=0$ under the assumption that the solution of the arising KdV equation is the exact two-soliton solution.
%The first two equations in \eqref{KdV system q} determine a dynamical system for the motion of the vortex in the "regular fluid particle" approximation.} 

The exact two-soliton solution of the KdV equation \eqref{KdV 0}
 (see,  for example, \cite{Johnson})  has the following form:
\begin{equation}\label{KdVsol2}
\hspace{-0.2cm} \mathfrak{u}(x,t)=\frac{4 c h^2}{3 }\Big\{\Big[\log\Big(2\cosh[\mathcal{E}_-(x,t)]+2\mathcal{K}\cosh\left[\mathcal{E}_+(x,t)+\log\mathcal{K}\right]\Big)\Big]_x\Big\}_x
\end{equation}
with the notations:
\begin{align}
\begin{split}
\mathcal{E}_1(x,t):=&K_1\left(x-x_{0,1}-\left(1+\frac{2}{3}h^2K_1^2\right)\, ct\right)\\
\mathcal{E}_2(x,t):=&K_2\left(x-x_{0,2}-\left(1+\frac{2}{3}h^2K_2^2\right)\, ct\right)\\
\mathcal{E}_+(x,t):=&\mathcal{E}_1(x,t)+\mathcal{E}_2(x,t)\\
\mathcal{E}_-(x,t):=&\mathcal{E}_1(x,t)-\mathcal{E}_2(x,t)\\
\mathcal{K}:=&\frac{K_2-K_1}{K_2+K_1}
\end{split}
\end{align}
where $x_{0,1}$, $x_{0,2}$, $K_1$, $K_2$ are constants such that $\mathcal{K}>0$. \\
For the solution \eqref{KdVsol2}, the two equations \eqref{KdV system q free} which determine the trajectory of the vortex in the ``regular fluid particle" approximation} can be written as
\begin{align}\label{75}
 \begin{split}
& \dot{q_1}=\frac{4 c h^2}{3 }\Big\{\Big[\log\Big(2\cosh[\mathcal{E}_-(q_1,t)]+2\mathcal{K}\cosh\left[\mathcal{E}_+(q_1,t)+\log\mathcal{K}\right]\Big)\Big]_{q_1}\Big\}_{q_1} \\
&
 \dot{q_2} =  (q_2+h)\frac{4 c h^2}{3 }\Big\{\Big[\log\Big(2\cosh[\mathcal{E}_-(q_1,t)]+2\mathcal{K}\cosh\left[\mathcal{E}_+(q_1,t)+\log\mathcal{K}\right]\Big)\Big]_{q_1}\Big\}_{q_1q_1}
\end{split}
\end{align}
The next order approximation includes terms  as in \eqref{q system epsilon}, where $\mathfrak{u}(x,t)$ is now  given by \eqref{KdVsol2}.  

The first equation in the system \eqref{75} depends only on $q_1(t)$ and $t$. Although it is more difficult to integrate this equation, we again conclude that the trajectories are bounded in a finite interval. The first observation is that, at the leading order, the time derivative is  $-c$ multiplied by the derivative with respect to $x$.
Since  $\dot{q_1}$ is of order $\epsilon,$ and $(1+\frac{2}{3}h^2K_1^2)\, c$, $(1+\frac{2}{3}h^2K_2^2)\, c$, are of order 1, it follows that $q_1(t)-x_{0,1}-(1+\frac{2}{3}h^2K_1^2)\, ct \to \mp \infty$, $q_1(t)-x_{0,2}-(1+\frac{2}{3}h^2K_2^2)\, ct \to \mp \infty$ as $t \to \pm \infty$. We also have\\ $\Big[\log\Big(2\cosh[\mathcal{E}_-(q_1,t)]+2\mathcal{K}\cosh\left[\mathcal{E}_+(q_1,t)+\log\mathcal{K}\right]\Big)\Big]_{q_1} \to \mp (K_1+K_2)$  as $t \to \pm \infty$. Therefore,
\begin{align}\label{range2}
q_1(\infty)-q_1(-\infty)\approx \frac{8h^2}{3}(K_1+K_2).
\end{align}
As a vertical shift estimation, the expression \eqref{vert_shift} can be used, where $K=\text{max}(K_1,K_2)$ is the bigger soliton parameter.

\begin{figure}[h!]
\begin{center}
(i$_a$)\fbox{\includegraphics[totalheight=0.16\textheight]{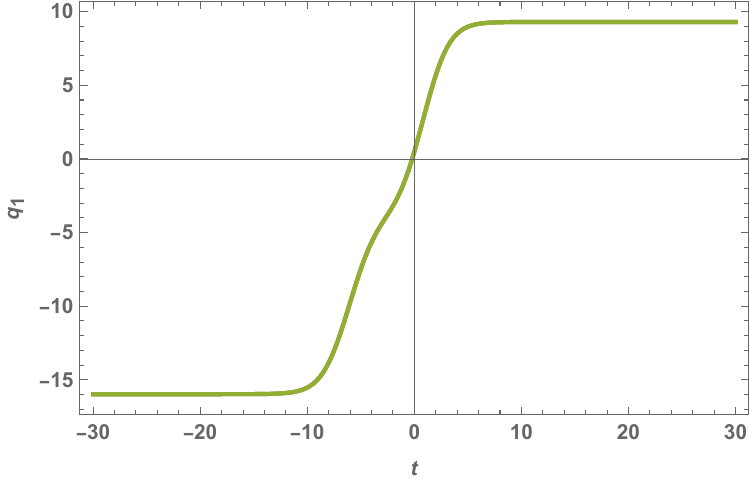}}
(i$_b$)\fbox{\includegraphics[totalheight=0.16\textheight]{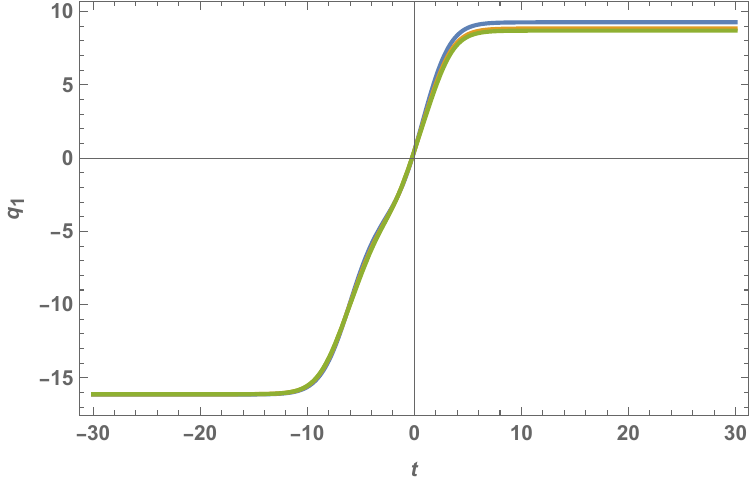}}
(ii$_a$)\fbox{\includegraphics[totalheight=0.16\textheight]{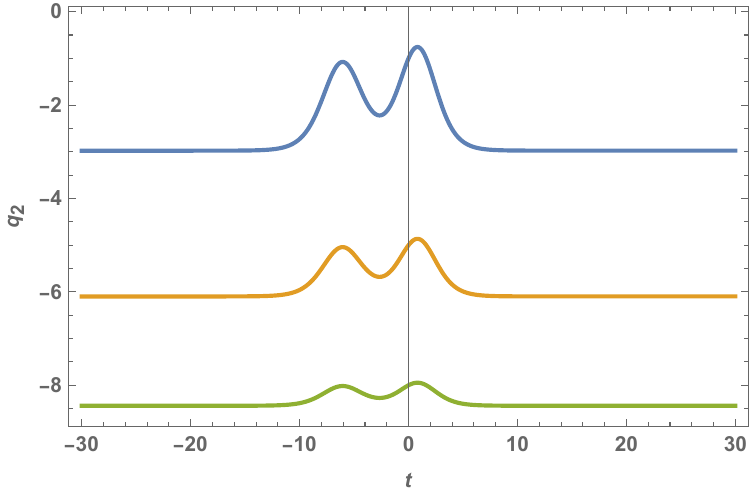}}
(ii$_b$)\fbox{\includegraphics[totalheight=0.16\textheight]{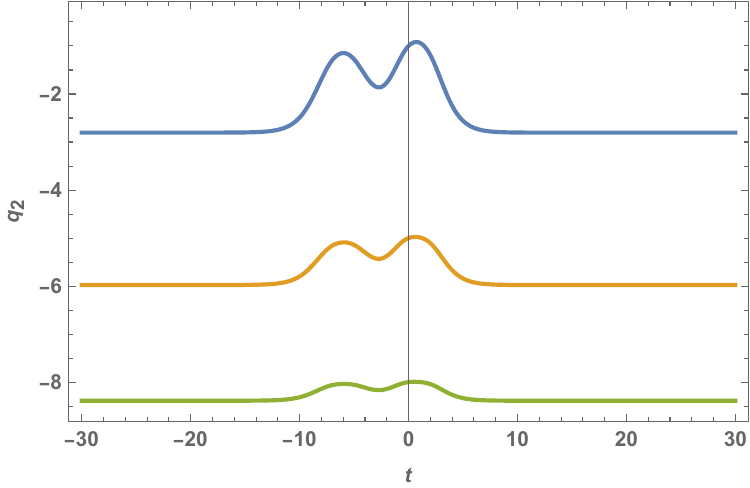}}
(iii$_a$)\fbox{\includegraphics[totalheight=0.09\textheight]{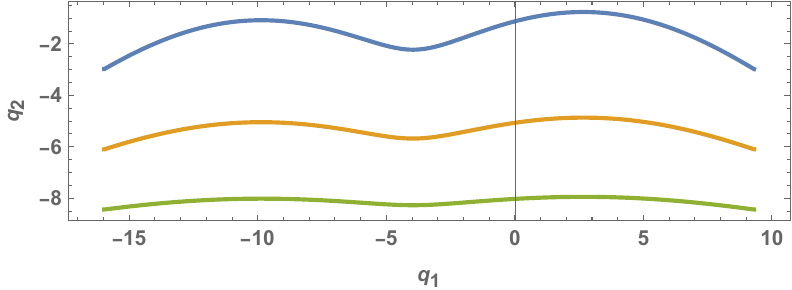}}
(iii$_b$)\fbox{\includegraphics[totalheight=0.09\textheight]{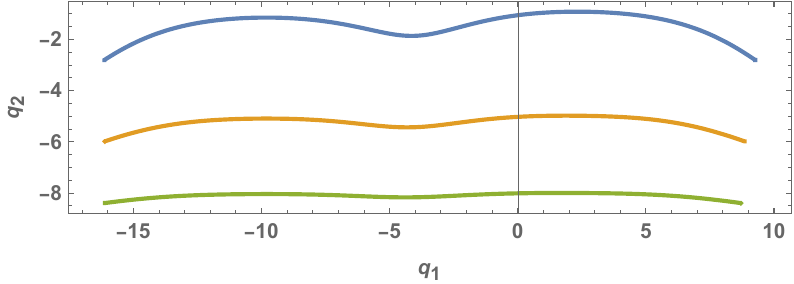}}
\caption{(i) The evolution in time of $q_1$; (ii) The evolution in time of $q_2$;
(iii) Vortex trajectories at different depth. 
The panels labeled with sub-index $a$ are for the system \eqref{75}, the panels labeled with index $b$ are for the system \eqref{q system epsilon} with the higher-order terms,  $\mathfrak{u}(x,t)$  given by \eqref{KdVsol2}. The parameter values are $x_{0,1}=0$, $x_{0,2}=0$, $h=10$m,   $K_1=0.04$m$^{-1}$,  $K_2=0.05$ m$^{-1}$,  the initial condition $q_1(0)=0.5$  and the initial condition $q_2$ takes different values: $q_2(0)=-1$ (blue), $q_2(0)=-5$ (orange), $q_2(0)=-8$ (green).}
\label{fig:KdV2soliton-1}
\end{center}
\end{figure}

\begin{figure}[h!]
\begin{center}
(i$_a$)\fbox{\includegraphics[totalheight=0.16\textheight]{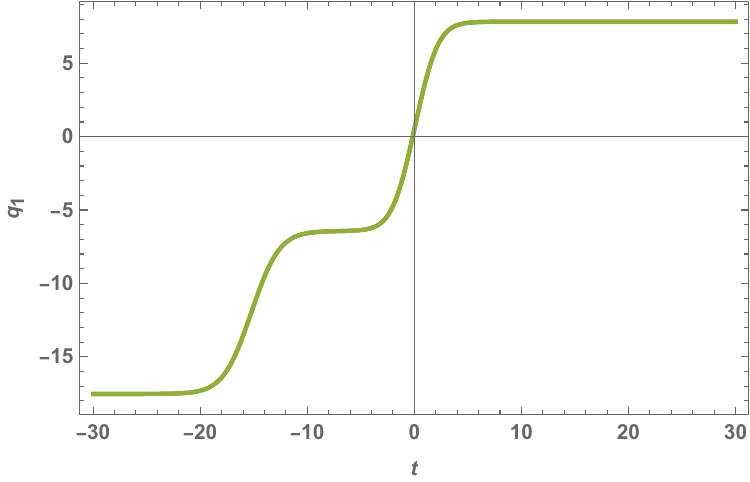}}
(i$_b$)\fbox{\includegraphics[totalheight=0.16\textheight]{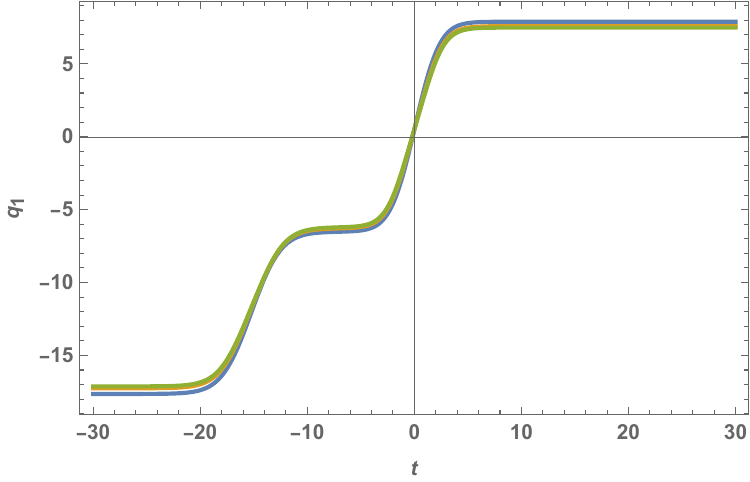}}
(ii$_a$)\fbox{\includegraphics[totalheight=0.16\textheight]{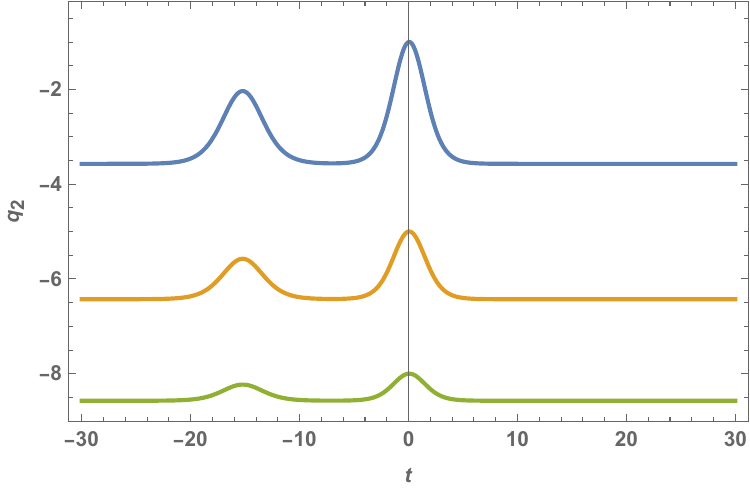}}
(ii$_b$)\fbox{\includegraphics[totalheight=0.16\textheight]{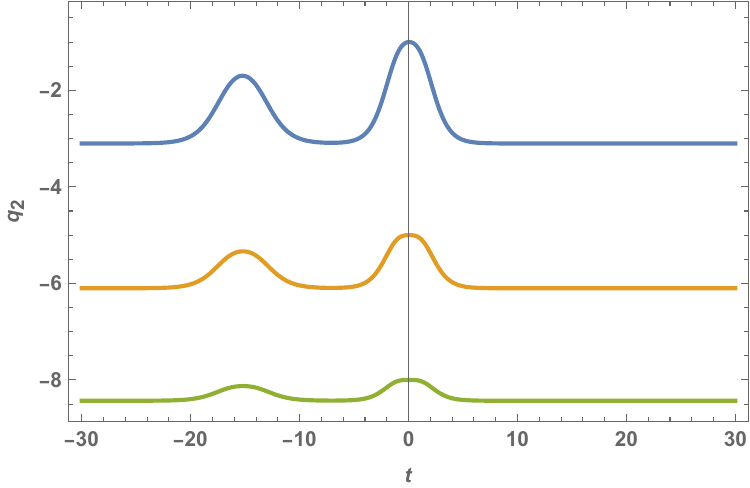}}
(iii$_a$)\fbox{\includegraphics[totalheight=0.089\textheight]{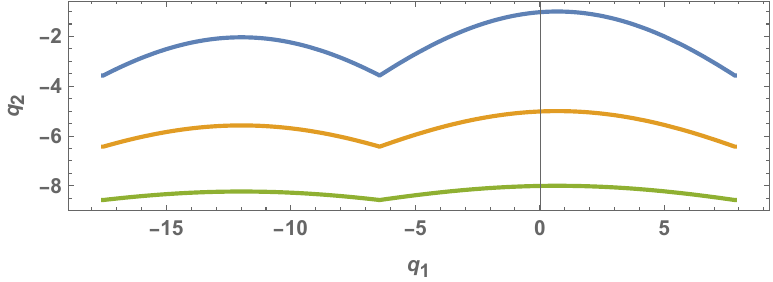}}(iii$_b$)\fbox{\includegraphics[totalheight=0.089\textheight]{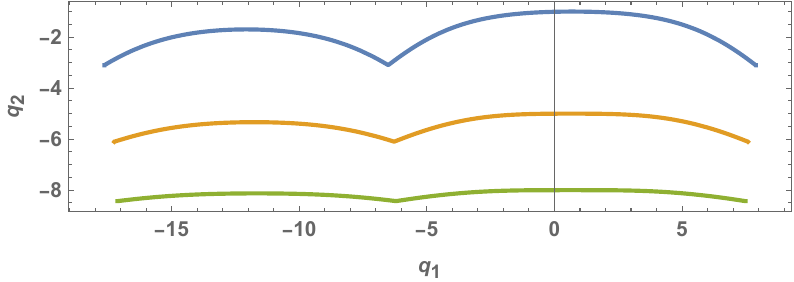}}
\caption{(i) The time evolution of $q_1$; (ii) The time evolution of $q_2$;
(iii) Vortex trajectories at different depth.
The panels labeled with sub-index $a$ are plots for the system \eqref{75}, those with sub-index $b$ are plots for the system \eqref{q system epsilon} with the higher-order terms,  $\mathfrak{u}(x,t)$  given by \eqref{KdVsol2}. The parameter values are  $x_{0,1}=100$, $x_{0,2}=0$, $h=10$m,   $K_1=0.04$m$^{-1}$,   $K_2=0.05$ m$^{-1}$,  the initial condition $q_1(0)=0.5$  and the initial condition $q_2$ takes different values: $q_2(0)=-1$ (blue), $q_2(0)=-5$ (orange), $q_2(0)=-8$ (green).}
\label{fig:KdV2soliton-2}
\end{center}
\end{figure}

The graph of $q_1$ versus $t$, the graph of $q_2$ versus $t$, and the graph of $q_2$ versus $q_1$ is plotted in Fig. \ref{fig:KdV2soliton-1},   Fig. \ref{fig:KdV2soliton-2}, for the following parameter values:
  $h=10$m,  $g=9.81$m/s$^2$, $K_1=0.04$m$^{-1}$, $K_2=0.05$m$^{-1}$, the initial condition $q_1(0)=0.5$  and  the initial condition for $q_2$   takes different values:  $q_2(0)=-1$ (blue), $q_2(0)=-5$ (orange), $q_2(0)=-8$ (green). The panels with the index $a$ are for the system \eqref{75}, the panels with the index $b$ are for the system \eqref{q system epsilon} with the higher-order terms,  $\mathfrak{u}(x,t)$  given by \eqref{KdVsol2}.
The differences between    Fig. \ref{fig:KdV2soliton-1} and Fig. \ref{fig:KdV2soliton-2} are the values of $x_{0,1}$ and $x_{0,2}$: these are  $x_{0,1}=0$ and $x_{0,2}=0$ in  Fig. \ref{fig:KdV2soliton-1}, that is, at $t=0$ the two solitons interact,  and  $x_{0,1}=100$ and $x_{0,2}=0$  in  Fig. \ref{fig:KdV2soliton-2}, that is, at $t=0$ the  two solitons are far away from each other. The trajectories are extended for both positive and negative $t.$
We observe in Fig. \ref{fig:KdV2soliton-1} ($iii_a$) and Fig. \ref{fig:KdV2soliton-2} ($iii_a$) that the point vortex trajectories are bounded and the range of the interval is always in agreement with \eqref{range2}, that is, $q_1(\infty)-q_1(-\infty)\approx  \frac{800}{3}\cdot0.09=24.$\\
We point out that, as in Section \ref{4.1}, the differences between the outputs in the panels with label a and those with label b are extremely insignificant, and  the shape of the trajectories and their boundedness are also preserved.

\section{Numerical investigation of the KdV equation perturbed by one vortex}\label{numerics}

We consider the  coupled system  \eqref{KdV system q}
with the initial conditions
\begin{align} \label{initial soliton}
\begin{split}
& q_1(0)=q_{1,0} \\
&  q_2(0) =q_{2,0}\\
& \mathfrak{u}(x, 0)=\frac{4 c h^2 K^2}{3 \cosh^2[K(x-x_0)]},
\end{split}
\end{align}
where the soliton parameters, that is,  its position  $x_0$ at $t=0$  and $K$ as well as the initial coordinates of the vortex $q_{1,0},$ $q_{2,0}$ are constants. The elevation of the surface is calculated from \eqref{eta final}.  The details of the numerical scheme are given in Appendix \ref{ApA}. 

In what follows, we consider $h=10$m, $g=9.81$m/s$^2$, $x_0=0$ and we take different values for $K$, $q_1(0)$, $q_2(0)$ and the vortex strength $\omega^*$.

\subsection{Soliton-vortex interactions}
In the previous sections, the initial conditions at $t=0$ correspond to the instant when the crest of the soliton is exactly above the vortex. However, for numerical modeling, this choice is very inconvenient, since it would require that the initial conditions be imposed in the middle of the soliton-vortex interaction, which is the main object of the investigation. Therefore, in the numerical setup, we consider a soliton that travels some distance before reaching the vortex location. This distance is chosen long enough so that the vortex does not "feel" the soliton initially, and short enough so that the vortex does not move far from its specified initial position.

\begin{figure}[h!]
(a)\fbox{\includegraphics[totalheight=0.16\textheight]{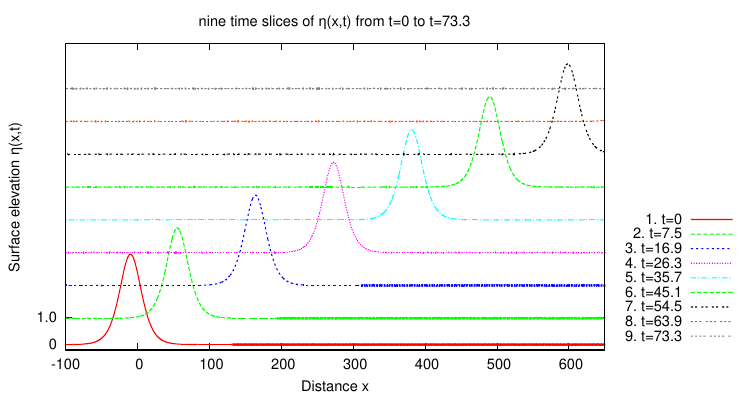} }
(b)\fbox{\includegraphics[totalheight=0.16\textheight]{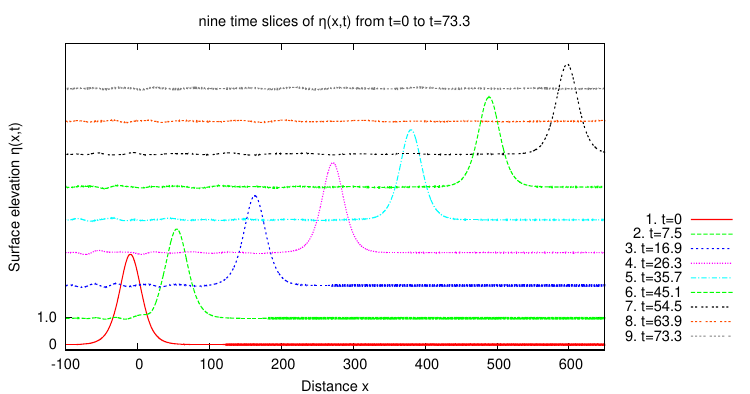}} 
(c)\fbox{\includegraphics[totalheight=0.16\textheight]{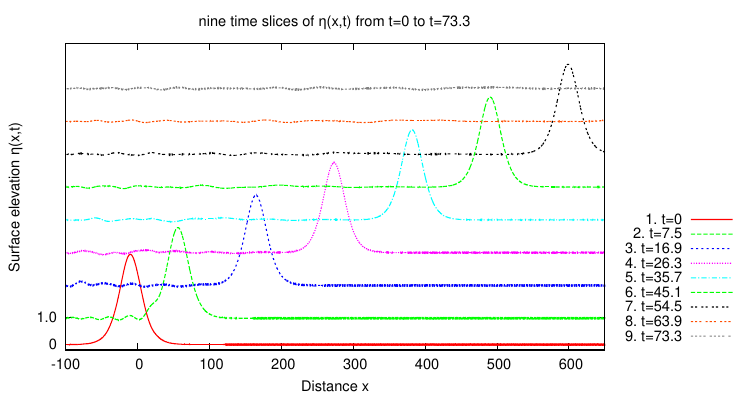}} 
(d)\fbox{\includegraphics[totalheight=0.16\textheight]{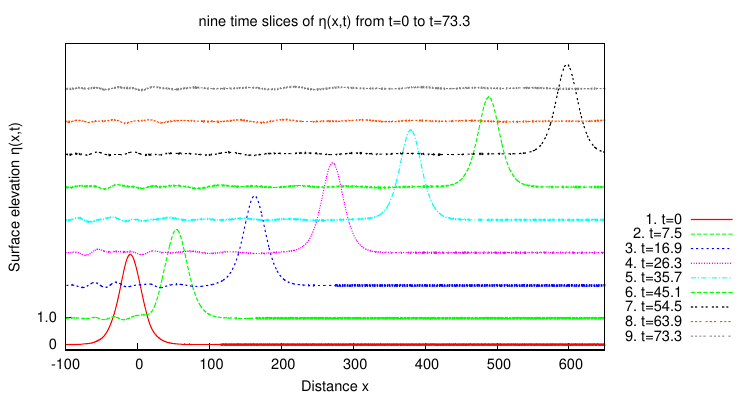}} 
(e)\fbox{\includegraphics[totalheight=0.16\textheight]{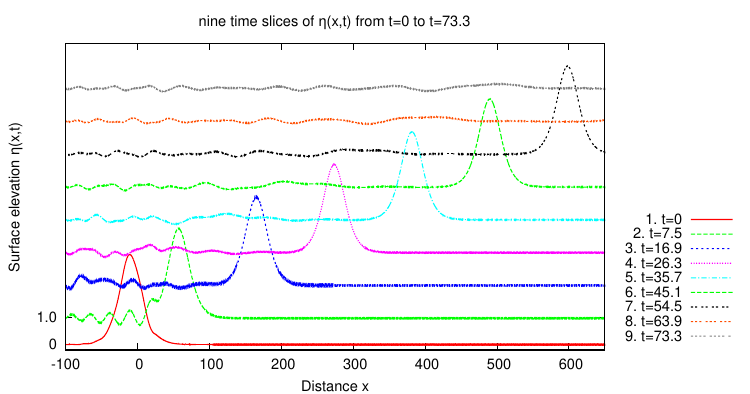}} 
(f)\fbox{\includegraphics[totalheight=0.16\textheight]{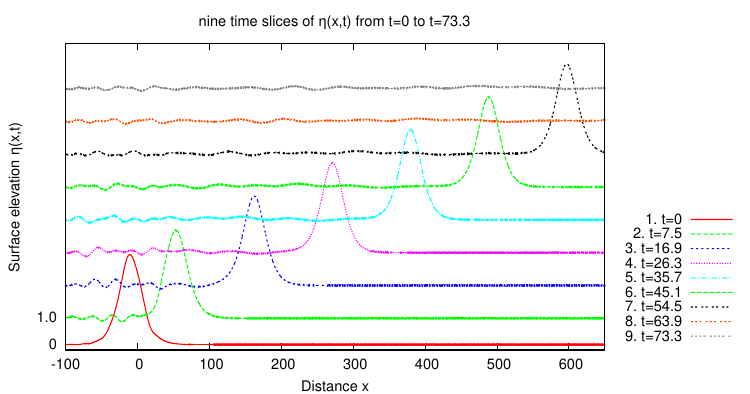}} 
(g)\fbox{\includegraphics[totalheight=0.16\textheight]{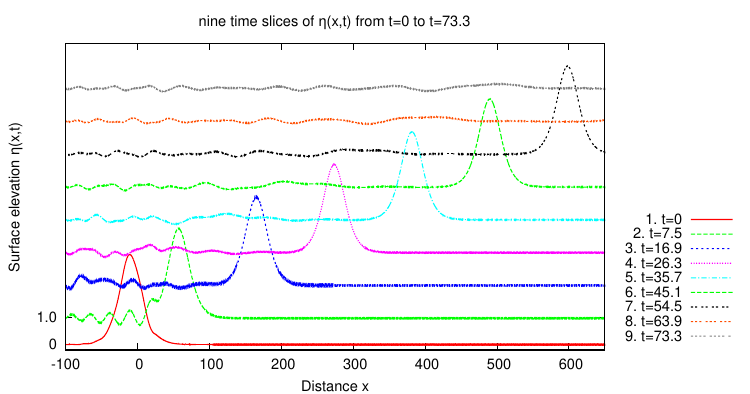}} 
 \caption{
 The surface motion \eqref{eta final} of the soliton-vortex interaction. The the initial conditions are KdV soliton with $x_0=0$, $K=0.05$ m$^{-1}$ and the vortex, situated at  $q_1(0)=25$, $q_2(0)=-5$.  
 The vortex strength has the values:
(a) $\omega^*=0$; (b) $\omega^*=20$; (c) $\omega^*=-20$; (d) $\omega^*=30$; (e)  $\omega^*=-30$; (f) $\omega^*=50$; (g) $\omega^*=-50$.} \label{Figure6}
  \end{figure}

\begin{figure}
 (a) \includegraphics[width=5in]{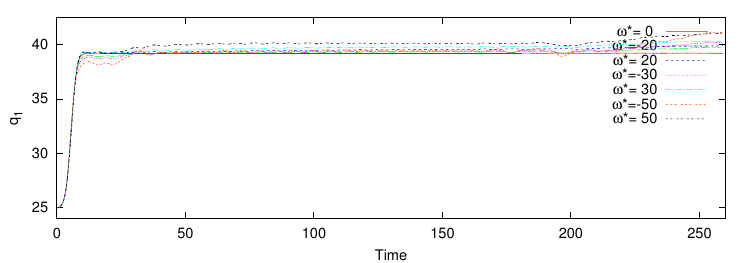}\\
 (b)  \includegraphics[width=5in]{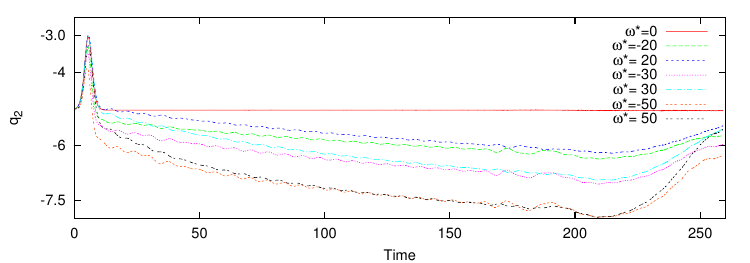}\\
 (c) \includegraphics[width=5in]{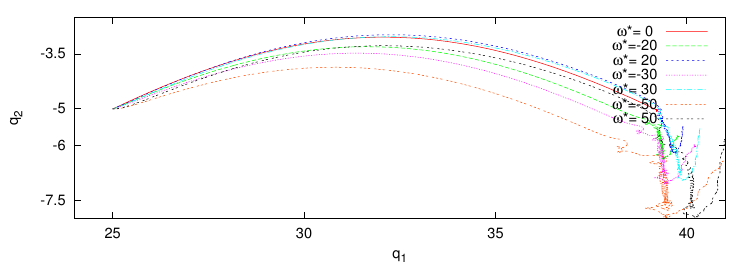}
  \caption{ (a) The evolution in time of $q_1$; (b) The evolution in time of $q_2$; (c) The  trajectories of the point  vortex. In all pictures,  the initial KdV soliton is located at $x_0=0$, $K=0.05$m$^{-1}$,  and the vortex is initially at $q_1(0)=25$, $q_2(0)=-5$.  The results for different values of the vorticity, that is, $\omega^*=0,\pm20,\pm30,\pm 50$, are drawn using different colours, as mentioned in each figure.}\label{Figure7}
  \end{figure}

\begin{figure}
(a) \includegraphics[width=5in]{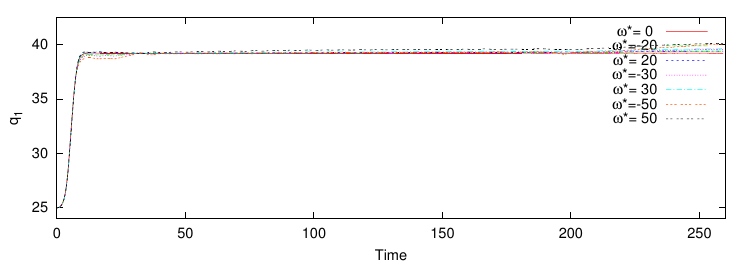}\\
(b) \includegraphics[width=5in]{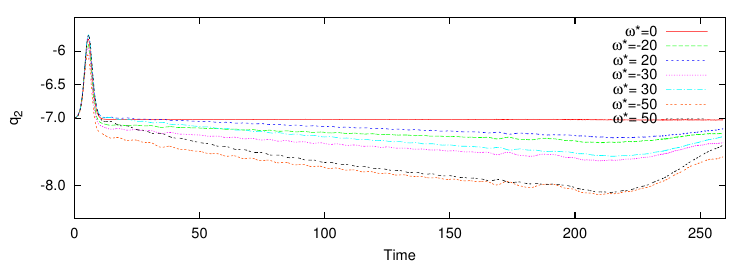}\\
(c)  \includegraphics[width=5in]{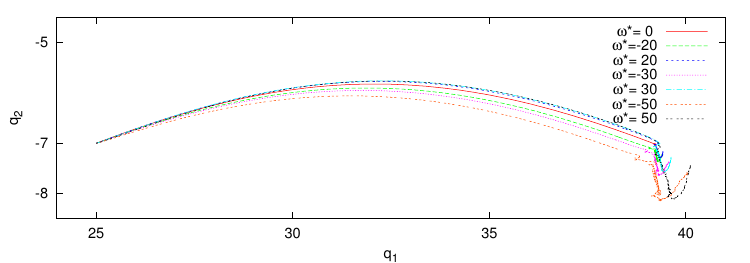}
\caption{ (a) The evolution in time of $q_1$; (b) The evolution in time of $q_2$; (c) The  trajectories of the point  vortex. In all pictures,  the initial KdV soliton is located at $x_0=0$, $K=0.05$m$^{-1}$,  and the vortex is initially at $q_1(0)=25$, $q_2(0)=-7$.  The results for different values of the vorticity, that is, $\omega^*=0,\pm20,\pm30,\pm 50$, are drawn using different colours, as mentioned in each figure.}\label{Figure8}
  \end{figure}

In Fig. \ref{Figure6}, we present the surface motion as a result of the soliton-vortex interaction between the soliton, initially defined at $x=x_0=0$ in \eqref{initial soliton} with $K=0.05$ m$^{-1}$ and the vortex, initially at $q_1(0)=25$, $q_2(0)=-5,$ for different values of the vortex strength, $\omega^*=0$, $\omega^*=\pm 20$, $\omega^*=\pm 30$, and $\omega^*=\pm 50$. The effect on the surface is in the form of some very small radiation waves, which travel to the left and disappear with time, as could be seen in Fig. \ref{Figure6}(b) -- Fig. \ref{Figure6}(g). These waves are most pronounced for $\omega^*=-50$, see Fig. \ref{Figure6}(g). We observe that the soliton is very stable and completely unaffected by these radiation waves. \\
In Fig. \ref{Figure7},  we plot the graph of $q_1$ versus $t$, the graph of $q_2$ versus $t$ and the graph of $q_2$ versus $q_1$, for all the values of the vorticity mentioned above. By comparing Fig. \ref{Figure7} and Fig.  \ref{fig:different_q2}, we observe  that the output when $\omega^*=0$, that is, when the vortex is just a regular fluid particle,  coincides with the  approximation from the previous Section \ref{4.1},  when the vortex  moves in the velocity field created by an unperturbed KdV soliton. The point vortex trajectories are bounded and the length of the horizontal displacement is approximately the same as estimated  in  Section \ref{4.1}, that is, $\approx 14.3$ m. We also observe that, for small enough values of $\omega^*$,  the situation is qualitatively similar to the one considered in Section \ref{4.1}, the ``regular fluid particle" approximation. 
For larger values of the vortex strength, i.e. $\omega^*=-30$ or $\omega^*=-50$, we observe more significant deviations from this approximation. This was to be expected given the fact that the unperturbed KdV soliton approximation is derived for a small values of vorticity strengths $\frac{| \omega^* |}{h \sqrt{gh}} \approx \frac{| \omega^*| }{100}  \ll 1.$\\
 When the vortex is initially placed deeper, for example, at $q_2(0)=-7$ instead of $q_2(0)=-5$,  the interaction with the free surface is weaker and the radiation waves are smaller. This can be seen in Fig. \ref{Figure8}, where except for the value of  the initial depth of the vortex,  all  other parameter values are the same as those used for the plots in Fig. \ref{Figure7}.
We also notice that, the vortex motion is again close to the ``regular fluid particle" approximation, for a range of vortex strengths.

\subsection{Large time - large distance effects of the surface-vortex interactions}

\begin{figure}
 (a) \includegraphics[width=5in]{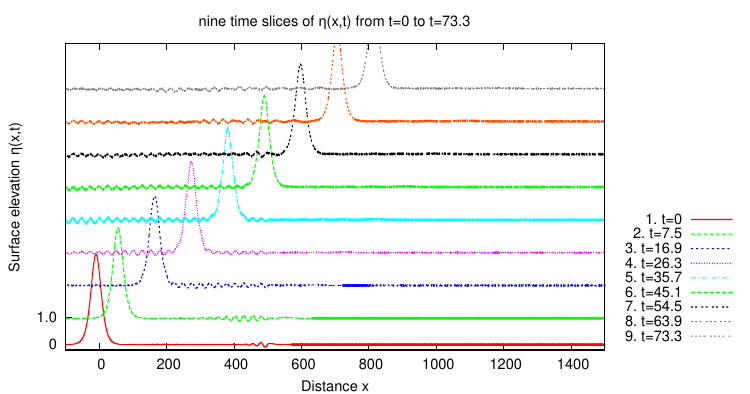}\\
(b)  \includegraphics[width=5in]{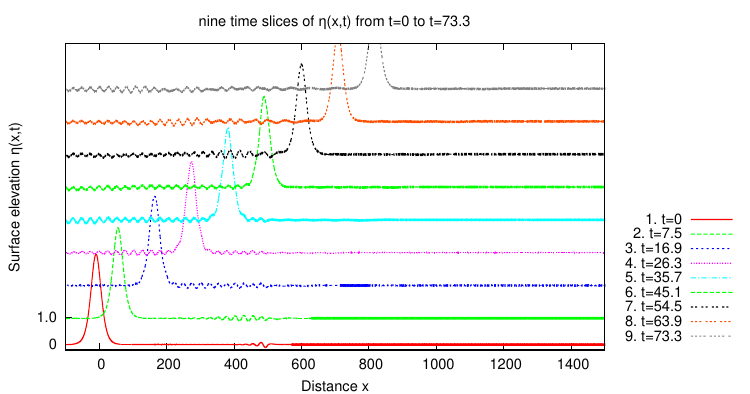}
  \caption{
 The surface motion \eqref{eta final} of the soliton-vortex interaction, with the initial KdV soliton $x_0=0$, $K=0.05$ m$^{-1}.$ The vortex  situated initially at  $q_1(0)=500$, $q_2(0)=-5$.
  The vortex strength has the values:
(a) $\omega^*=50$; (b) $\omega^*=-50$. }
\label{Figure9}
\end{figure}

Let us now suppose that the soliton is initially located  far away from the vortex, for example, if again  $x_0=0$ but instead of $q_1(0)=25$ we choose $q_1(0)=500$.
In this case, it takes a significant amount of time for the soliton to reach out the location of the vortex.
During this time, the vortex is affected by the soliton and the free surface motion in general.
The main features of these interactions are illustrated in Fig. \ref{Figure9}; we present only the cases with $\omega^*=\pm 50$, where the results are very pronounced.
By comparing the outputs in Fig. \ref{Figure9} with the ones in Fig. \ref{Figure6},  obtained for a vortex initially situated at $q_1(0)=25$, much closer to the soliton, we observe much bigger and stronger radiation waves on the surface. This is due to the larger time for their generation, related to the longer traveling time of the soliton to the vortex location.

\begin{figure}
(a)   \includegraphics[width=5in]{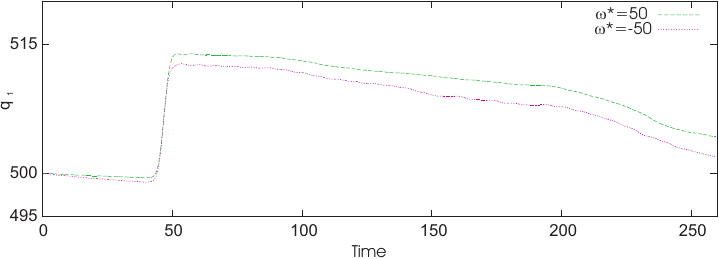}\\
(b)   \includegraphics[width=5in]{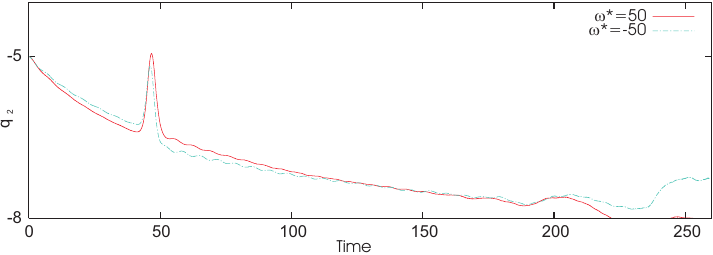}\\
(c)  \includegraphics[width=5in]{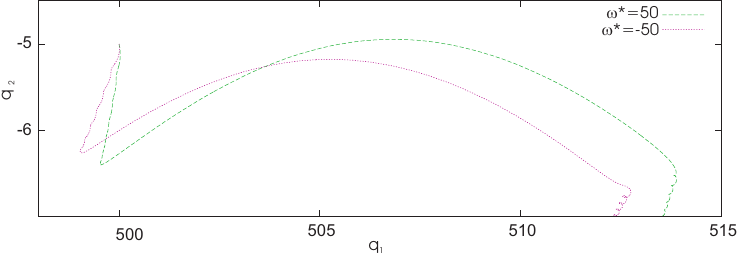}
  \caption{
(a) The evolution in time of $q_1$; (b) The evolution in time of $q_2$; (c) The  trajectories of the point  vortex. In all pictures,  the initial KdV soliton is located at $x_0=0$, $K=0.05$m$^{-1}$,  and the vortex is initially at $q_1(0)=500$, $q_2(0)=-5$.  The results for different  values of the vorticity, that is, $\omega^*=\pm 50$, are drawn using different colours, as mentioned in each figure.}\label{Figure10}
\end{figure}

\begin{figure}
 \includegraphics[width=5in]{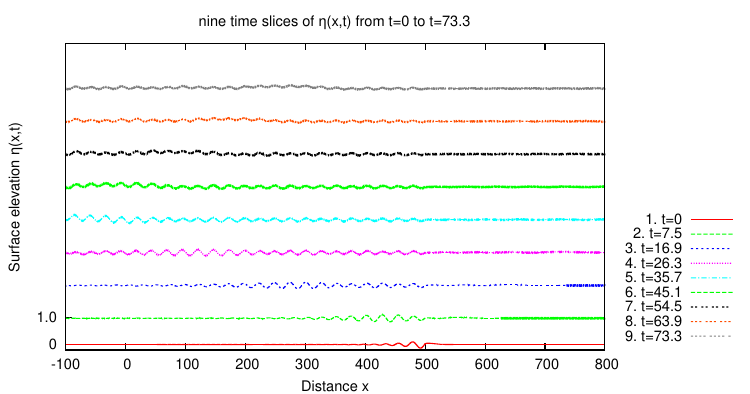}
  \caption{The surface motion, $\eta(x,t)$  as in \eqref{eta final} with no soliton ($x_0=0$, $K=0$) and the vortex  situated initially at  $q_1(0)=500$, $q_2(0)=-2.5$. The vortex strength is $\omega^*=30$. The $x$-coordinate is along the horizontal axis.}\label{Figure11}
\end{figure}

The soliton is again stable and not significantly affected by the radiation waves. We also observe that, during the longer traveling time, the vortex is moving slowly towards the bottom, see Fig. \ref{Figure10}. \\
Finally, if there is no soliton, that is, $x_0=0$, $K=0$, the vortex is generating only radiation waves on the surface, Fig. \ref{Figure11}.

\begin{figure}
 \includegraphics[width=5in]{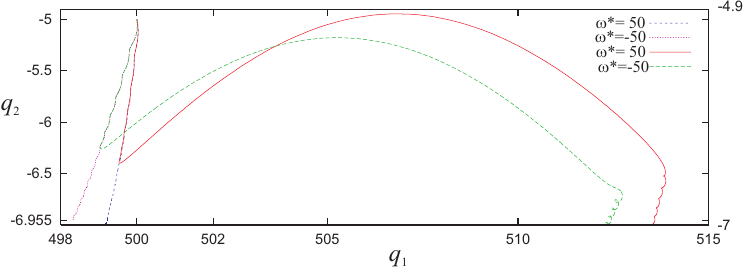}
  \caption{Comparison to the motion in Fig. \ref{Figure10} (c) (in green and red colours) and a free sinking of the vortex in the absence of soliton (in blue and lilac colors).  The vortex is situated initially at  $q_1(0)=500$, $q_2(0)=-5.0$. The vortex strengths are $\omega^*=\pm 50$.}\label{Figure10d}
\end{figure}

Finally, we make some comments on the apparent irregular  trajectories of the point vortex  on the right side of Fig. \ref{Figure7} (c) and Fig. \ref{Figure8} (c).
We observe the following common feature: after the interaction with the soliton, the vortex continues to sink down to the bottom for some time, with some perturbations (wavy ripples) due to the radiation waves on the surface, then will move up. Looking at Fig. \ref{Figure10} (c) for example, the pattern is more evident: initially, the soliton starts from a point located far away from the location of the vortex, and until it arrives, the vortex sinks down and slightly to the left; then the soliton kicks in, it shifts the vortex to the right, and finally, when the soliton is gone, the vortex keeps sinking down to the bottom for some time, with some apparent ripples due to the radiation waves on the surface.
To better emphasize this behavior, we  look at the point vortex trajectories in the case without soliton. We draw these paths  in Fig. \ref{Figure10d}, for two different values of the vortex strength, i.e., $\omega^*=\pm 50$. Starting initially from the point $(q_1=500, q_2=-5),$ the vortex is sinking down and slightly to the left; see the trajectories in blue and lilac colors in Fig. \ref{Figure10d}.  For a better comparison, we also keep the trajectories in red and green colors as in Fig. \ref{Figure10} (c).

\subsection{Stability of the soliton}

\begin{figure}
(a) \includegraphics[width=5in]{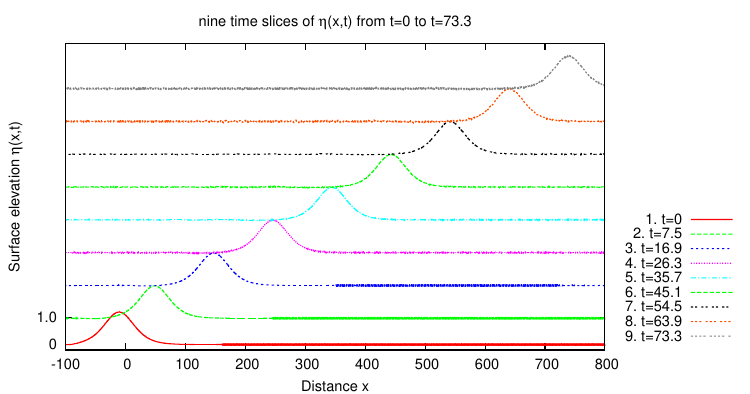}\\
(b) \includegraphics[width=5in]{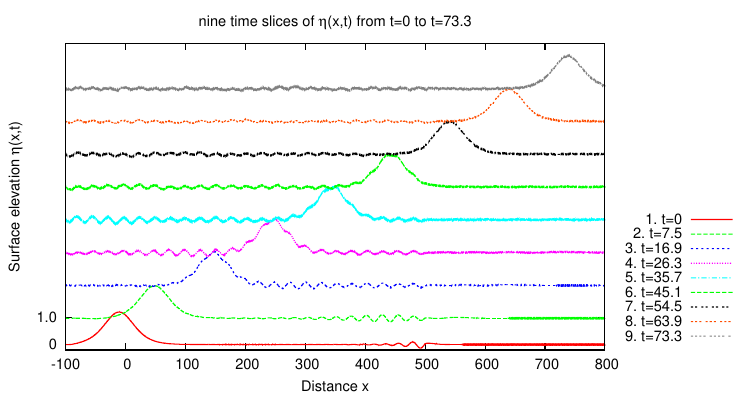}
  \caption{
The surface motion \eqref{eta final} due to the soliton-vortex interaction, with the initial KdV soliton $x_0=0$, $K=0.03$ m$^{-1}$ and the vortex  situated initially at  $q_1(0)=500$, $q_2(0)=-2.5$.  
The vortex strength has the values: (a) $\omega^*=0$;  (b) $\omega^*=30$.}    \label{Figure12}
\end{figure}

The numerical simulations show that the soliton is extremely robust and stable and cannot be destroyed in the course of its interaction with the vortex.
The motion of the soliton in the absence of the vortex is presented in Fig. \ref{Figure12}(a).
The initial position of the soliton is $x_0=0$ and the soliton parameter is $K=0.03$ m$^{-1}$, that is, the soliton is small compared to those considered in the previous examples. By placing a point vortex at $q_1(0)=500$, $q_2(0)=-2.5$, we see in  Fig. \ref{Figure12}(b) how the small amplitude soliton interacts with the strong radiation waves created by the vortex. However, the soliton remains intact and fully recovers after the interaction.

\section{Conclusions}
\begin{figure}
\includegraphics[width=5in]{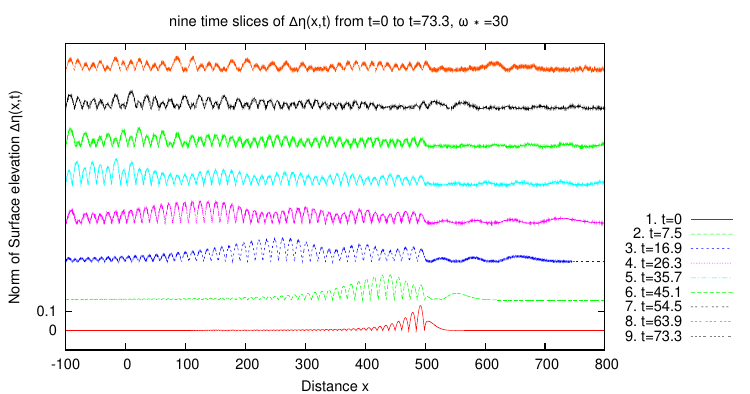}\\
  \caption{The difference  \eqref{Delta_eta} between the outputs in Fig. \ref{Figure12} (b) and Fig. \ref{Figure12} (a).  }    \label{Figure16}
\end{figure}

In the adopted framework (small soliton amplitudes and small vortex strength), we have established that the system of interacting vortex and soliton could be well described on a qualitative level by the decoupled system \eqref{KdV 0} with either \eqref{KdV system q free} or \eqref{q system epsilon} with surface elevation \eqref{eta 68}. 
This ``regular fluid particle" approximation reflects the weak influence of the vortex on the soliton. 
        As an explicit measure of this influence, we examined the difference between $\eta_{30}(x,t),$ the surface deviation when $\omega ^*=30$ and $\eta_{0}(x,t),$  the one when $\omega ^*=0$  (which is the exact KdV soliton in the absence of vortex) 
\begin{equation}\label{Delta_eta}
    \Delta \eta (x,t)=|\eta_{30}(x,t) - \eta_0(x,t)|
\end{equation}
in the case where the parameters are chosen as in Fig. \ref{Figure12}.  The results are presented in Fig. \ref{Figure16}.
We observe that the soliton is  ``invisible" in this figure, therefore the difference is of the magnitude of the ``ripples" due to the radiation waves. Again, this confirms that the soliton is practically not affected by the vortex. The radiation waves are created almost entirely by the vortex (located at $q_1(0)=500$), and in addition the soliton interacts elastically with the radiation waves, recovering completely its original shape far from the vortex - as seen in Fig. \ref{Figure12} (b).

Therefore, the approximation suggested is far simpler than the coupled system \eqref{KdV system q}, in essence it is a dynamical system, given the fact that the KdV model is exactly solvable and its stable solutions are the soliton solutions. This will allow for the study of other vortex configurations involving several vortices moving under solitons propagating on the surface. The results will be reported in a forthcoming publication.

In this study, we have considered only idealized point vortices and have not discussed the mechanism of vortex formation, which is a fundamental research problem in fluid dynamics. A possible mechanism of vortex creation is the interaction of the flow created by a surface wave with the bottom topography; see, for example, \cite{Ch} and the references therein.   In \cite{Gorban} a numerical study of the viscous interaction between a non-breaking solitary wave and a submerged step on the bottom is presented. The formation of two large-scale opposite vortices shedding from the step edge is detected in both upstream and downstream directions. For experimental purposes, the vortices can be created by a pair of vertically oriented counter-rotating flaps \cite{Fish, Will}.

\appendix
\section{Finite-difference implementation for the KdV equation with a vortex perturbation}\label{ApA}

We need to numerically solve the equation \eqref{KdV system q} with respect to the functions $q_1(t)$, $q_2(t)$ and $\mathfrak{u}=\mathfrak{u}(x,t)$
\begin{align}
{\dot q_1}&=\mathfrak{u}(q_1(t),t),  \label{eq1}\\
{\dot q_2}&=-(q_2+h)\mathfrak{u}_x(q_1(t),t), \label{eq2}\\
\mathfrak{u}_t+c\mathfrak{u}_x+\frac{gh^2}{6}\mathfrak{u}_{xxx}&+\left[\frac{1}{4}(\mathfrak{u}-T_1)^2+\frac{1}{2}\mathfrak{u}^2\right]_x+T_2=0\label{eq3}\end{align}
where $$T_1= \frac{\omega^*}{2h}\,.\,\frac{\sin\frac{\pi q_2}{h}}{\cosh\frac{\pi(x-q_1)}{h}-\cos\frac{\pi q_2 }{h}},$$
\begin{align*}T_2=&-\frac{\pi \omega*}{4h^2}\left[\frac{\sin\frac{\pi q_2}{h}\sinh\frac{\pi(x-q_1)}{h}}{\left(\cosh\frac{\pi(x-q_1)}{h}-\cos\frac{\pi q_2 }{h}\right)^2} \mathfrak{u}(q_1)\right. \\&\hspace{2cm}\left.+(q_2+h)\frac{1-\cos\frac{\pi q_2}{h}\cosh\frac{\pi(x-q_1)}{h}}{\left(\cosh\frac{\pi(x-q_1)}{h}-\cos\frac{\pi q_2}{h} \right)^2}\mathfrak{u}_x(q_1)\right].\end{align*}

We assume the initial conditions for the vortex to be constants $$q_1(0)=q_{1,0},\quad q_2(0)=q_{2,0}$$ and for the KdV equation to be in the form of the one-soliton solution, as in \eqref{initial soliton}, $$\mathfrak{u}(x,t=0)=\frac{4ch^2K^2}{3\cosh^2[K(x-x_0)]}.$$

In the $x-$direction we consider a uniform mesh in the interval  $[-L_1, L_2],$  $x_i = (i - 1)\Delta x$ with a spatial step $\Delta x = (L_1 + L_2)/(N - 1)$, where $N$ is the total number of grid points in the interval. The discretization of time is $t^n = n\Delta t$, where $\Delta t$ is the increment in time.

Consequently, $\mathfrak{u}_i^n$ and $\mathfrak{u}_i^{n+1}$ denote the values of the function $u$ on the \textit{i}-th spatial point and time stages $t^n$ and $t^{n+1}$. We note that $\omega^*=0$ reduces the system to the unperturbed KdV equation; see, for example, \eqref{eq3}. For the unperturbed KdV equation we can construct a nonlinear difference scheme that is convergent,
\begin{multline}\label{eq4} 
\frac{\mathfrak{u}_i^{n+1}-\mathfrak{u}_i^n}{\Delta t} + \frac{c}{4\Delta x}(\mathfrak{u}_{i+1}^{n+1}-\mathfrak{u}_{i-1}^{n+1}+\mathfrak{u}_{i+1}^n-\mathfrak{u}_{i-1}^n)\\
+\frac{g h^2}{24(\Delta x)^3} (-\mathfrak{u}_{i-2}^{n+1}+2 \mathfrak{u}_{i-1}^{n+1}-2 \mathfrak{u}_{i+1}^{n+1} + \mathfrak{u}_{i+2}^{n+1} -\mathfrak{u}_{i-2}^n+2\mathfrak{u}_{i-1}^n-2\mathfrak{u}_{i+1}^n + \mathfrak{u}_{i+2}^n)\\
+\frac{3}{16(\Delta x)}\left[(\mathfrak{u}_{i+1}^{n+1})^2-(\mathfrak{u}_{i-1}^{n+1})^2 + (\mathfrak{u}_{i+1}^n)^2-(\mathfrak{u}_{i-1}^n)^2\right]=0.
\end{multline}

Such a scheme will be stable with respect to the physical time-stepping if the iterative procedure with respect to the nonlinear terms is convergent.

Since the scheme (\ref{eq4}) cannot be implemented directly, because it is nonlinear, we follow the idea of \cite{CDM} to introduce internal iterations, namely
\begin{multline}\label{eq5}
\frac{\mathfrak{u}_i^{n+1,k+1}-\mathfrak{u}_i^n}{\Delta t}+ \frac{c}{4\Delta x}\left(\mathfrak{u}_{i+1}^{n+1,k+1}-\mathfrak{u}_{i-1}^{n+1,k+1}+\mathfrak{u}_{i+1}^n- \mathfrak{u}_{i-1}^n\right)\\
+\frac{g h^2}{24(\Delta x)^3} (-\mathfrak{u}_{i-2}^{n+1,k+1}+2 \mathfrak{u}_{i-1}^{n+1,k+1}-2 \mathfrak{u}_{i+1}^{n+1,k+1} + \mathfrak{u}_{i+2}^{n+1,k+1}\\ -\mathfrak{u}_{i-2}^n+2 \mathfrak{u}_{i-1}^n-2\mathfrak{u}_{i+1}^n + \mathfrak{u}_{i+2}^n)\\
+\frac{3}{16(\Delta x)}\left[\mathfrak{u}_{i+1}^{n+1,k+1}\mathfrak{u}_{i+1}^{n+1,k}- \mathfrak{u}_{i-1}^{n+1,k+1}\mathfrak{u}_{i-1}^{n+1,k} + (\mathfrak{u}_{i+1}^n)^2-(\mathfrak{u}_{i-1}^n)^2\right]=0.
\end{multline}

In this way, for the current iteration of the unknown function (superscript $n + 1, k +1$), we have an implicit system with five-diagonal band matrix. We begin from an initial condition $\mathfrak{u}^{n+1,0}=\mathfrak{u}^n$ and carry out the internal iterations (repeating the calculations for the same time step $(n + 1)$ with an increasing value of the superscript $k$) until convergence.

Note that the initial condition is a very good guess that is within $O(\Delta t)$ of the sought solution for $\mathfrak{u}$. This makes the convergence of the internal iterations very fast. We have performed the due numerical experiments to verify this fact. Even
for very large values of the time increment $\Delta t$ we did not encounter instability of the internal iterations. In each case under consideration, we selected $\Delta t$ such that no more than six internal iterations were required to reach the precision of $10^{-12}$.
After the internal iterations converge, the solution of the non-linear scheme is obtained by setting $\mathfrak{u}^{n+1} \equiv \mathfrak{u}^{n+1,k+1}$. In such a way, we have a fully implicit, nonlinear, and conservative scheme. 

If $\omega^* \ne 0$ two more terms need to be added to the KdV equation, from \eqref{eq3} these are, $$\frac{1}{4}\left(-2 \mathfrak{u} T_1+ T_1^2\right)_x+T_2,$$ and, respectively the following terms need to be added to \eqref{eq5}
\begin{align*}&-\frac{1}{8\Delta x}\left[T_{1,i+1}\left(\mathfrak{u}_{i+1}^{n+1,k+1}+\mathfrak{u}_{i+1}^n\right)-T_{1,i-1}\left(\mathfrak{u}_{i-1}^{n+1,k+1}+\mathfrak{u}_{i-1}^n\right)\right]\\
&\hspace{2cm}+\frac{1}{8\Delta x}\left(T_{1,i+1}^2-T_{1,i-1}^2\right)+T_{2,i}.
\end{align*} 

For the inversion of the five-diagonal $N\times N$ matrix we use a generalized algorithm based on Gaussian elimination with pivoting (for details, see \cite{read, TodChri13, marrak16}). 
The scheme was thoroughly validated through standard numerical tests that involve halving the space and time increments.
The global truncation error of the time approximation was verified by the Runge principle and confirmed to the second order of accuracy in time. In a similar fashion, we found that the global spatial truncation error is also second-order, $O((\Delta x)^2)$.

In each time stage $n$ prior to solving \eqref{eq1}, \eqref{eq2} we determine where the quantity $q_1$ is located. If $x_{i-1}\le q_1 \le x_i$, $i=1,N$ then we set $$ q_1=x^*=k_1x_{i-1} +k_2 x_i,\quad \mathfrak{u}^*=k_1u_{i-1} +k_2 u_i\,$$ where 
$$k_1=\frac{x^*-x_{i-1}}{\Delta x},\,\, k_2=\frac{x_i-x^*}{\Delta x},\,\, \Delta x=x_i-x_{i-1}.$$
Finally, from \begin{align*}\frac{q_1^{n+1}-q_1^n}{\Delta t}=& \mathfrak{u}(q_1^n=x^*,t)\equiv \mathfrak{u}^* \\
\frac{q_2^{n+1}-q_2^n}{\Delta t}=&-(q_2^n+h)\mathfrak{u}_x(q_1^n=x^*,t)\equiv -(q_2^n+h)\mathfrak{u}^*_x\end{align*}
we obtain the difference formulae for the vortex coordinates for the new time stage, $n+1,$
\begin{align*}
 q_1^{n+1}&= \mathfrak{u}^* \Delta t  + q_1^n,    \\
 q_2^{n+1}&=q_2^n -  (q_2^n+h)\mathfrak{u}^*_x \Delta t.
\end{align*}

{\small
\subsection*{Acknowledgements}

D.I.-K. and R.I. would like to thank the Erwin Schr\"odinger International Institute for Mathematics and Physics
(ESI), Vienna (Austria) for the opportunity to participate in the Research in Teams project: Variational approaches to modelling geophysical waves and flows (July 1, 2022 — July 31, 2022) where this work has been initiated.
 The authors thank three anonymous referees for their important critical remarks and suggestions. 

\subsection*{Funding}
 The research carried out by D.I.-K. is financially supported by a grant of the Ministry of Research, Innovation and Digitization, CCCDI-UEFISCDI,  project number ROSUA-2024-0001, within PNCDI IV.
 The research carried out by R.I. is financially supported by {\it Research Ireland} under Grant number 21/FFP-A/9150. 

\subsection*{Conflicts of interest/Competing interests} 

The authors have no conflicts of interest to declare that are relevant to the content of this article.

\subsection*{Data Availability Statement} 

Data sharing is not applicable to this article, as no data sets were generated or analyzed during the current study.  

}

\end{document}